\documentclass[notitlepage,nofootinbib,preprintnumbers,amssymb,superscriptaddress]{revtex4-1}

\usepackage{amsfonts,amssymb,amsmath,graphicx,color,bm}
\definecolor{ultramarine}{rgb}{0.07, 0.04, 0.56}
\definecolor{cadmiumgreen}{rgb}{0.0, 0.42, 0.24}
\definecolor{indigo(dye)}{rgb}{0.0, 0.25, 0.42}
\usepackage[linktocpage=true]{hyperref}
\hypersetup{
colorlinks=true,
citecolor=ultramarine,
linkcolor=cadmiumgreen,
urlcolor=indigo(dye),
}

\newcommand{\f}[2]{\frac{#1}{#2}}  
\newcommand{\mk}[1]{\left( #1 \right)}  
\newcommand{\kk}[1]{\left[ #1 \right]}  
  
\newcommand{\be}{\begin{equation}}  
\newcommand{\ee}{\end{equation}}
\newcommand{\bem}{\begin{bmatrix}}
\newcommand{\eem}{\end{bmatrix}}
\newcommand{\Mpl}{M_{\rm Pl}}
\newcommand{\e}{\epsilon}
\newcommand{\E}{\mathcal{E}}

\newcommand{\pa}{\partial}
\newcommand{\vx}{{\boldsymbol x}}

\newcommand{\ti}{\tilde}
\newcommand{\da}{\dagger}
\newcommand{\bS}{\mathbb{S}}
\newcommand{\bV}{\mathbb{V}}
\newcommand{\bT}{\mathbb{T}}
\newcommand{\Lap}{\bigtriangleup}

\begin{document}

\preprint{RESCEU-26/16}

\title{  
Fundamental theorem on gauge fixing at the action level
}

\author{Hayato Motohashi}
\affiliation{Kavli Institute for Cosmological Physics, 
The University of Chicago, Chicago, Illinois 60637, USA}
\author{Teruaki Suyama}
\affiliation{Research Center for the Early Universe (RESCEU),
Graduate School of Science, The University of Tokyo, Tokyo 113-0033, Japan}
\author{Kazufumi Takahashi}
\affiliation{Research Center for the Early Universe (RESCEU),
Graduate School of Science, The University of Tokyo, Tokyo 113-0033, Japan}
\affiliation{Department of Physics, Graduate School of Science, 
The University of Tokyo, Tokyo 113-0033, Japan}

\begin{abstract}

Regardless of the long history of gauge theories, it is not well recognized under which condition gauge fixing at the action level is legitimate.
We address this issue from the Lagrangian point of view, and prove the following theorem on the relation between gauge fixing and Euler-Lagrange equations:
In any gauge theory, if a gauge fixing is complete, i.e., the gauge functions are determined uniquely by the gauge conditions, 
the Euler-Lagrange equations derived from the gauge-fixed action are equivalent to those derived from the original action supplemented with the gauge conditions.
Otherwise, it is not appropriate to impose the gauge conditions before deriving Euler-Lagrange equations as it may in general lead to inconsistent results.
The criterion to check whether a gauge fixing is complete or not is further investigated.
We also provide applications of the theorem to scalar-tensor theories and make comments on recent relevant papers on theories of modified gravity, in which there are confusions on gauge fixing and counting physical degrees of freedom.

\end{abstract}
\maketitle  


\section{Introduction}
\label{sec:int} 

Symmetry plays a fundamental role in physics.
In particular, gauge theories are constructed up on some gauge symmetry or gauge transformation, under which the Lagrangian is invariant up to total derivative.
It implies that the theory has redundant degrees of freedom (DOFs), which make the analysis of gauge theories involved.
To cope with this difficulty, the method of so-called gauge fixing has been employed in many situations, such as electrodynamics, general relativity and theories of modified gravity.
Gauge fixing helps us eliminate the gauge DOFs and extract physical DOFs.

In implementing a gauge fixing in the dynamical analysis of gauge theories, there are two options commonly used:
(i)~The first option is to fix the gauge after deriving the equations of motion (EOMs) for all the fundamental variables appearing in the original action. 
In this case, we have all the EOMs corresponding to all the variables.
Since the general solutions of the EOMs contain arbitrary functions corresponding to the gauge DOFs and they can be eliminated by fixing the gauge completely, the solutions of the EOMs supplemented with complete gauge-fixing conditions should correctly represent the physical DOFs.
(ii)~The second option is to fix the gauge at the action level, and then derive the EOMs from the gauge-fixed action. 
This process often significantly simplifies the derivation of EOMs and subsequent dynamical analysis thanks to less number of variables that one must manage.
However, since gauge fixing reduces the number of independent fundamental variables, the number of EOMs obtained by variational principle in this approach is apparently less than the number of fundamental variables.
Then, one may naively wonder if imposing gauge-fixing conditions at the action level changes the dynamical properties of the system and the conclusions drawn from it would be incorrect.
At the same time, one may also naively think that gauge fixing at the action level, if it is complete, should lead to correct conclusions since the gauge-fixed action contains only physical DOFs without missing any physical information.

Along this line, \cite{henneaux1992quantization,Garriga:1997wz} discussed the validity of complete gauge fixing in the language of Hamiltonian mechanics.
Although their results imply that complete gauge fixing at the action level is harmless, the aforementioned problem on the loss of EOMs in the Lagrangian formalism has not been clarified.
\cite{Lagos:2013aua} dealt with this issue, but their arguments are restricted to the case of cosmological perturbation theory and justification for more generic cases has not been established.

Another motivation of this work, which may be related to the incomplete understanding of the role of gauge fixing at the action level in the Lagrangian formalism, is the following confusions in recent works, in particular, on counting the number of physical DOFs of theories of modified gravity.
For example, in the context of de Rham-Gabadadze-Tolley (dRGT) massive gravity~\cite{deRham:2010ik,deRham:2010kj}, the dynamics of the St\"{u}ckelberg fluctuations around Minkowski background was investigated in \cite{Koyama:2011wx} imposing a gauge condition at the action level, and arrived at an inconsistent counting of physical DOFs, which is explained in \cite{Motloch:2014nwa}.
In addition, a class of isotropic self-accelerating solutions~\cite{Gratia:2012wt} in dRGT massive gravity is shown to have a different number of DOFs for isotropic perturbations in a special choice of coordinate~\cite{Khosravi:2013axa}, as the constant-time surface of such a coordinate system coincides with the characteristics of the isotropic perturbations~\cite{Motloch:2015gta}.

In light of this situation, it is worthwhile to make a clear statement on the relation between the EOMs derived from the gauge-fixed action and those derived from the original action.
In this paper, we prove that if a gauge fixing is complete, i.e., if the gauge functions are fixed without ambiguity of integration constant, then the EOMs that are lost by the gauge fixing can be recovered from the remaining components of the EOMs.
Conversely, if a gauge fixing is incomplete, it is not pertinent to impose it in the action before deriving the EOMs.
Indeed, the inconsistency in the DOF counting in \cite{Koyama:2011wx} originates from an incomplete gauge fixing in the action.
It seems somewhat obvious physically, however, to the best of our knowledge, no explicit and mathematically rigorous proof has been found before.

Before closing this section, we comment on the relevant works \cite{Pons:1995ss,Pons:2009ch}, in which the role of imposing (holonomic) gauge-fixing conditions in the action was investigated.
The author of these papers claimed that a gauge-fixed action provides the same set of equations as the original set of EOMs plus the gauge conditions, if the former is supplemented with the lost primary constraints corresponding to the lost gauge symmetries.
On the other hand, our work is specialized to the case of {\it complete} gauge fixing, and it gives a stronger result:~no additional constraint is needed in the analysis under a gauge-fixed action if the gauge fixing is complete.

This paper is organized as follows.
In \S \ref{ssec:toy}, we first provide a simple example that is useful to get an intuition of the main theorem.
In \S \ref{sec:fth}, we present a general proof of the theorem on complete gauge fixing at the action level for a general field theory with multiple fields and multiple gauge symmetries in arbitrary spacetime dimensions.
Then in \S \ref{sec:anm}, we focus on analytical mechanics as a special case of field theories, and explicitly obtain the criterion for complete gauge fixing by using some mathematical technique which is explained in Appendix~\ref{DAE}.
We provide applications of the theorem to scalar-tensor theories of gravity in \S \ref{sec:app}.
Furthermore, we make some comments on the conflicts in the above papers in \S \ref{comments}.
Finally, we draw our conclusions in \S \ref{conclusions}.

\section{Toy model}
\label{ssec:toy}

Before considering general gauge theories in \S \ref{sec:fth}, let us begin with a simple toy model to understand an essential point of the relation between gauge fixing and Euler-Lagrange equations.
Consider a Lagrangian consisting of two variables~$x(t)$ and $y(t)$,
	\be \label{toyL} L = \f{1}{2} (\dot x - \ddot y)^2. \ee
As it will soon turn out, this Lagrangian is equivalent to the one for a single point particle freely moving in one-dimensional space.
This Lagrangian is invariant under a gauge transformation 
	\be \label{toygauge} x \to x + \dot \xi , \quad y\to y + \xi ,  \ee
with $\xi(t)$ being an arbitrary function. 
Clearly, $y=0$ fixes $\xi$ without any ambiguity, $\xi=-y$, and hence it is a complete gauge-fixing condition.
On the other hand, $x=0$ does not fix $\xi$ completely as a constant DOF remains.
The EOMs for $x$ and $y$ without gauge fixing are given by
	\be 
	\label{toyeom} \E_x = -\ddot x + \dddot y=0,~~~\E_y = - \dddot x + y^{(4)}=0.  
	\ee
We can verify that these two equations are related as  
	\be - \dot \E_x + \E_y=0, \ee
which is the well-known Noether identity~\cite{Noether:1918zz} in this model.
It is clear that $\E_y=0$ is a redundant equation as it can be derived from a derivative of $\E_x=0$.  
In contrast, $\E_x=0$ is an independent equation in the sense that the remaining equation~$\E_y=0$ cannot recover it uniquely.

First, let us consider the situation in which one derives EOMs from \eqref{toyL} and then imposes the gauge condition~$y=0$ or $x=0$.  For the complete gauge-fixing condition~$y=0$, the result is given by
	\be
	\label{toyeomg1} -{\ddot x}=0,~~~-{\dddot x}=0,~~~y=0.
	\ee
The second equation is automatically satisfied by the first equation, and thus the basic equation for $x$ is a second-order differential equation.
Hence the solution describes the motion of a point particle with constant velocity, and we need two initial conditions to determine the time evolution of $x$.
Therefore, this system has one DOF.\footnote{The number of DOFs is defined by half of the number of initial conditions.}

For the incomplete gauge-fixing condition~$x=0$, one obtains
	\be
	\label{toyeomg2} \dddot y=0,~~~y^{(4)}=0,~~~x=0.
	\ee
Again the second equation is redundant, and the system is described by a third-order differential equation.  Thus one needs three initial conditions, one of which corresponds to a residual gauge DOF.  Indeed, by imposing an initial condition, e.g., $y(0)=0$ to fix the residual gauge DOF, the system requires two initial conditions, which is consistent with the result of the previous case.
Therefore, even in an incomplete gauge fixing, the analysis should be consistent so long as the gauge conditions are imposed after deriving EOMs and the residual gauge DOFs are eliminated by some additional conditions.

Second, let us consider imposing the gauge condition at the Lagrangian level.
For the complete gauge-fixing condition~$y=0$, the gauge-fixed Lagrangian is given by
	\be L = \f{1}{2} \dot x^2. \label{lagx} \ee
The resulting EOMs are
	\be
	{\ddot x}=0,~~~y=0.
	\ee
This set of equations is evidently the same as \eqref{toyeomg1} obtained by imposing the gauge condition at the EOM level.
Thus, we have explicitly verified in this simple model that the set of EOMs derived from the gauge-fixed Lagrangian does not lose any information if the gauge fixing is complete.

On the other hand, if one uses the incomplete gauge fixing by setting $x=0$, the Lagrangian reads 
	\be L = \f{1}{2} \ddot y^2. \label{lagy} \ee
Then the EOMs are given by 
	\be
	y^{(4)}=0,~~~x=0. \label{y4eom}
	\ee
This is clearly inconsistent with \eqref{toyeomg2}.
Since we need four initial conditions to determine the time evolution of $y$, the number of DOFs is two.
Furthermore, since the Lagrangian consists of a higher derivative term and is nondegenerate, one of the DOFs corresponds to an Ostrogradsky ghost~\cite{Woodard:2015zca}.  
Note that even if one supplements an initial condition for $y$ to fix the residual gauge DOF, the system still requires three initial conditions.  In this sense, the analysis is not consistent, and this inconsistency cannot be resolved even if we take into account the residual gauge DOF.
A general lesson that we can learn from this simple example is that an incomplete gauge fixing at the Lagrangian level generically leads to an insufficient set of EOMs, or an incorrect counting of DOFs which may even contain Ostrogradsky ghosts.

The essential difference of the $x=0$ gauge from the $y=0$ gauge is that the lost EOM for $x$ is an independent EOM, which cannot be reproduced from the other EOM.
Therefore, one should just avoid to use the incompletely gauge-fixed Lagrangian~\eqref{lagy}, or after deriving the EOMs~\eqref{y4eom} one should derive the lost EOM for $x$ from the original Lagrangian without gauge fixing, and then impose the gauge condition $x=0$.
The resultant set of EOMs is then the same as \eqref{toyeomg2} and there is no inconsistency.

Complementary to the gauge-fixed Lagrangian analysis above, let us remark that one can confirm without gauge fixing that this 
model has one healthy DOF by Hamiltonian analysis.
To see this, let us first transform the Lagrangian~\eqref{toyL} into the following equivalent form:
	\be L = -\f{1}{2} q^2 + q(\dot x - \ddot y) = -\f{1}{2} q^2 + q \dot x + \dot q \dot y , \ee 
where the first equality is justified by using the EOM for the auxiliary variable~$q$, $q=\dot x - \ddot y$, and the second equality is valid up to total derivative.  
The advantage of expressing $L$ in this way is that the new Lagrangian contains at most first time derivatives of the variables and we can perform Hamiltonian analysis in a standard manner.
Let $\pi_X$ be canonical momenta conjugate to $X=x,y,q$.
Because of the degeneracy of the kinetic matrix, there is one primary constraint~$\pi_x - q \approx 0$.
One can then verify that the consistency condition for the primary constraint yields the secondary constraint~$\pi_y \approx 0$ and no further constraints are required.
Since the Poisson bracket of the two constraints vanishes, they are first-class constraints as expected from the gauge symmetry. 
Thus, this system has $\frac{1}{2} (6-2\times 2)=1$~DOF.
This one DOF is healthy since the Hamiltonian evaluated on the constraint surface reads $H=q^2/2$ and thus bounded below.\footnote{In addition to the straightforward Hamiltonian analysis, it is actually immediate to see that the Lagrangian~\eqref{toyL} satisfies the so-called degeneracy condition~\cite{Motohashi:2016ftl}, under which a general Lagrangian containing higher-order derivatives is free from Ostrogradsky ghosts that originate from the higher derivative terms.}

The lesson from the toy model is that, while it is appropriate to impose complete or incomplete gauge-fixing conditions after deriving EOMs, it requires a special care to impose them at the action level before deriving EOMs as there is a crucial difference between complete and incomplete gauge-fixing conditions.
The complete gauge-fixing condition $y=0$ could be imposed at the Lagrangian level without any inconsistency, whereas the incomplete gauge-fixing condition $x=0$ led to the inconsistent set of EOMs when imposed at the Lagrangian level.
The prescription is that one should just avoid to use the incompletely gauge-fixed Lagrangian, or supplement the lost EOM for $x$ by deriving it from the original Lagrangian without gauge fixing.
In the next section, we prove that the difference exists for general gauge theories.

\section{Proof of the theorem}
\label{sec:fth} 

In this section, we prove the main theorem of the present paper.
We first set up our notations and derive a key identity in \S \ref{ssec:fthset}, and then prove the theorem in \S \ref{sec:mth}.  
We discuss further extension of the theorem in \S \ref{ssec:bou}.

\subsection{Setup}
\label{ssec:fthset}

Having captured the essence of the main theorem in \S \ref{ssec:toy}, let us now consider a general field theory defined by the Lagrangian~$L=L(\phi^i, \pa_\mu \phi^i, \pa_\mu \pa_\nu \phi^i, \cdots; x^\mu)$ with multiple fields~$\phi^i=\phi^i(x^\mu)$ in $D$-dimensional spacetime, which is invariant up to total derivative under a general gauge transformation
	\be \label{gaugetrnsf} \phi^i \to \phi^i + \Delta_\xi \phi^i ,  \ee
where $\Delta_\xi\phi^i$ depend on gauge functions~$\xi^I(x^\mu)$ and their derivatives.
Here, $i=1,\cdots, n$ labels the fields and $I=1,\cdots,m$ labels the gauge symmetries, with $m<n$.

In such a theory with gauge symmetries, there exists an identity between the EOMs, which is known as Noether's second theorem~\cite{Noether:1918zz}.
Actually, this identity plays a crucial role in the proof of the main theorem given in \S \ref{sec:mth}.
Let us consider an infinitesimal gauge transformation
	\be
	\phi^i\to\phi^i+\Delta_\e\phi^i, \label{infgt}
	\ee
where $\Delta_\e \phi^i$ are linearized as
	\be \label{fgaut} \Delta_\e \phi^i = \sum_{p=0}^k F^{i(p)}_I \pa_{(p)} \e^I .  \ee
Here, we suppress indices for $p$th-order coefficients and derivative as 
	\be
	\begin{split} 
	F^{i(p)}_I &\equiv F^{i\mu_1\cdots \mu_p}_I , \\
	\pa_{(p)} &\equiv \pa_{\mu_1}\cdots \pa_{\mu_p}.
	\end{split}
	\ee
Note that $F^{i(p)}_I$ are functions of the fields $\phi^i$ and their derivatives, and they can also depend explicitly on $x^\mu$.
In \eqref{fgaut}, $p=0$ term is understood as $F^{i}_I \e^I$ without derivative, and all the other terms with $p\geq 1$ have $p$th derivative of the infinitesimal gauge functions~$\e^I$.  
Although the summation over repeated indices is implicit in principle, we sometimes restore the summation symbol for some indices for clarity.
Since the action is invariant under the infinitesimal gauge transformation~\eqref{infgt}, we obtain
	\be \label{gaugein} 0 = \Delta_\e S = \int d^Dx\,\E_i \Delta_\e \phi^i . \ee
Here, $\E_i$ are the EOMs for $\phi^i$, i.e., the Euler-Lagrange equations derived by the variational principle:\footnote{Strictly speaking, it is the set of equations~$\E_i=0$, not $\E_i$ themselves, that should be called EOMs. Nevertheless, we refer to $\E_i$ as EOMs throughout this article, which will not cause any confusion.}
	\be \E_i \equiv 
	\f{\pa L}{\pa \phi^i} - \pa_\mu \mk{ \f{\pa L}{\pa (\pa_\mu \phi^i)}} + \pa_\mu \pa_\nu\mk{ \f{\pa L}{\pa (\pa_\mu \pa_\nu\phi^i)} } - \cdots. 
	\ee
Plugging \eqref{fgaut} into \eqref{gaugein} and integration by parts yield
	\be 0 = \int d^Dx \kk{ \sum_{p=0}^k (-1)^p \pa_{(p)} \left( \E_i F^{i(p)}_I \right) } \e^I. \ee
Since $\e^I$ are arbitrary functions, we obtain the following key identity (Noether identity) between the EOMs:
	\be \label{fiden} \sum_{p=0}^k (-1)^p \pa_{(p)} \left( \E_i F^{i(p)}_I \right) = 0, \ee
for $I=1,\cdots,m$.

Let us remark that one can verify that the gauge transformation of the EOMs~$\E_i$ can be written as a linear combination of $\E_i$ and their derivatives.
This means that, as it should be, if a configuration of $\phi^i$ satisfies the EOMs, then its gauge transformation~$\phi^i+\Delta_\e \phi^i$ also satisfies the same set of EOMs.

\subsection{Main theorem}
\label{sec:mth}

Let us consider gauge fixing of a general field theory.
As mentioned in \S \ref{sec:int}, there are two methods commonly used for this purpose.
The first option is to fix the gauge after deriving all the EOMs.
The second option is to fix the gauge at the action level, and then derive EOMs from the simplified action, in which we obtain only a part of all the EOMs.

To illustrate the above point, let us consider the case where one sets first $m_g\,(\leq m)$ fields to zero:
	\be \phi^i=0 ,~~~(i=1,\cdots, m_g), \ee
by choosing the gauge functions~$\xi^I\,(I=1,\cdots, m_g)$ such that they satisfy $\Delta_\xi \phi^i = - \phi^i$, which is a generalization of the situation considered in \S \ref{ssec:toy}.
If one imposes the gauge conditions at the action level, one does not obtain the EOMs~$\E_i$ for the first $m_g$ fields.  
Whether one loses information or not depends on whether the lost EOMs are independent or redundant.
By redundant EOMs, we mean equations that can be recovered by using the Noether identity~\eqref{fiden} with the remaining EOMs~$\E_i\,(i=m_g+1,\cdots, n)$ derived from the gauge-fixed action.
In this case, no information is lost and the subsequent dynamical analysis is justified.
In contrast, if the lost EOMs are independent, one cannot recover them and the subsequent dynamical analysis is in general inconsistent. 
We would like to avoid such a type of gauge fixing at the action level.
Actually, it is possible to discern the two cases by checking whether the gauge fixing is complete or not.  
For a general gauge fixing, we shall prove the following theorem: 

\vskip0.3cm
{\bf Theorem.} {\it
Let the Lagrangian~$L=L(\phi^i, \pa_\mu \phi^i, \pa_\mu \pa_\nu \phi^i, \cdots;x^\mu)$ be invariant up to total derivative under a transformation~\eqref{gaugetrnsf}.
Consider imposing gauge conditions
	\be 
	\label{gengfix} f^I\left(\phi^i,\pa_\mu\phi^i,\pa_\mu\pa_\nu\phi^i,\cdots, \pa_{(\ell)} \phi^i;x^\mu \right) = 0 ,~~~(I=1,\cdots, m_g),
	\ee
in the action with $m_g\leq m$. If the gauge fixing is complete, i.e., the conditions~\eqref{gengfix} uniquely fix $m_g$ components of the gauge functions~$\xi^I$, then the $m_g$ components of the EOMs~$\E_i$ that are lost by imposing the gauge conditions at the action level can be recovered from the remaining $n-m_g$ components of $\E_i$.
}
\vskip0.3cm

Here, the term ``complete gauge fixing'' does {\it not} mean $m_g=m$, which uses all the gauge DOFs.  
Rather, a gauge fixing is defined to be complete when $m_g$ out of $m$ gauge functions~$\xi^I$ are determined without ambiguity of integration constant.  
One could consider a complete gauge fixing with $m_g<m$ (see the end of \S \ref{ssec:grp} and \S \ref{ssec:ssspert} for specific examples).
Since such a partial gauge fixing does not affect the rest gauge DOFs, we do not need the knowledge of all the gauge symmetries of a given theory.
In what follows, we simply denote by $\xi^I$ only the $m_g$ relevant components of the gauge functions.

One should also note that the gauge transformation must be defined so that the number of derivatives must be minimized.
Otherwise, one may misclassify a complete gauge fixing as an incomplete one.
To see this point, suppose a given theory is invariant under a gauge transformation of the form~\eqref{gaugetrnsf} and gauge conditions~$f^I=0$ fix the gauge functions~$\xi^I$ completely.
Now let us consider another transformation by $\zeta^I$:
	\be
	\phi^i\to\phi^i+\Delta_{\dot{\zeta}}\phi^i, \label{zetadot}
	\ee
which is obtained by the replacement $\xi^I\to\dot{\zeta}^I$ in \eqref{gaugetrnsf}, and thus the action is also invariant under \eqref{zetadot}.
For this new gauge transformation, the same set of gauge conditions~$f^I=0$ does not fix $\zeta^I$ uniquely since there remains ambiguity of functions that are constant in time, and it seems as if the gauge fixing is incomplete.
This misclassification originates from an inappropriate choice of the generators of the gauge transformation.
Though this example seems somewhat ridiculous, such a situation could arise in practice (see \S \ref{ssec:sss}).

To prove the Theorem, let us formulate the definition for the gauge fixing by \eqref{gengfix} to be complete in more convenient manner.
Suppose one could find gauge functions~$\xi^I$ that transform a given configuration of $\phi^i$ so that it satisfies the gauge conditions $f^I=0$.
Now let us consider an infinitesimal gauge transformation $\phi^i\to\phi^i+\Delta_\e \phi^i$ from such a configuration. 
If the gauge fixing is complete, there is no gauge transformation for which the transformed variables still satisfy the gauge conditions, namely, any infinitesimal gauge transformation spoils $f^I=0$.
This means that for complete gauge fixing the change of the gauge-fixing functions vanishes, $\Delta_\e f^I=0$, if and only if $\e^I=0$.
Obviously, $\e^I=0$ is always a solution for $\Delta_\e f^I=0$, but the point is that $\e^I=0$ is the unique solution.
The explicit form of $\Delta_\e f^I$ is given by
	\begin{align} 
	\Delta_\e f^I &= \sum_{q=0}^\ell \f{\pa f^I}{\pa\left(\pa_{(q)} \phi^i\right)} \pa_{(q)} (\Delta_\e \phi^i) \notag\\
	&= \sum_{p=0}^k \sum_{q=0}^\ell \f{\pa f^I}{\pa\left(\pa_{(q)} \phi^i\right)} \pa_{(q)} \mk{ F^{i(p)}_J \pa_{(p)} \e^J } \notag\\
	&\equiv \hat P^I_J \e^J,
	\label{Deltaf} 
	\end{align}
where $\hat P^I_J$ is an $m_g\times m_g$ matrix whose arguments are derivative operators acting on $\e^J$.
One can also express $\hat{P}^I_J$ in a simpler form.
With the aid of the Leibniz rule, one can sort out $\hat{P}^I_J$ by the order of derivative as
	\be
	\hat{P}^I_J=\sum_{p=0}^k \sum_{q=0}^\ell \sum_{r=0}^q \left(\begin{array}{c}
	q\\ r
	\end{array}\right)
	\left[\f{\pa f^I}{\pa\left(\pa_{(q)} \phi^i\right)}\pa_{(q-r)}F^{i(p)}_J\right]\pa_{(p+r)} \equiv\sum_{s=0}^{k+\ell} M^{(s)}{}^I_J \pa_{(s)}, \label{psort}
	\ee
where the explicit form of $M^{(s)}{}^I_J$ is
	\be
	M^{(s)}{}^I_J=\sum_{p={\rm max} \{ s-\ell,0 \}}^{{\rm min} \{ s,k \}}
	\sum_{q=s-p}^\ell \left(\begin{array}{c}
	q\\ s-p
	\end{array}\right)
	u_i^{I(q)} \pa_{(p+q-s)} F^{i(p)}_J,~~~
	u_i^{I(q)} \equiv \f{\pa f^I}{\pa\left(\pa_{(q)} \phi^i\right)}. \label{eq-Ms}
	\ee
To reiterate, the gauge fixing is complete if and only if
	\be
	\hat{P}^I_J\e^J=0 \label{PDAE2}
	\ee
is uniquely solvable.

A set of equations of this type is known as partial differential-algebraic equations (partial DAEs, or PDAEs).
Since $\hat{P}^I_J$ is a differential operator, at first sight, one may expect that the PDAE system~\eqref{PDAE2} is uniquely solvable if and only if all the differential parts in $\hat{P}^I_J$ vanish, namely, 
	\be \label{mg1cond} \hat{P}^I_J=M^{(0)}{}^I_J,~~~\left(\det M^{(0)}{}^I_J\ne 0\right). \ee  
This is indeed a necessary and sufficient condition for $m_g=1$, but is not necessary for $m_g \ge 2$.
For instance, for $m_g=2$, let us consider a case such that
	\be \label{exmg2}
	M^{(0)}=\left(
	\begin{array}{cc}
	a & 0 \\
	b & c \\
	\end{array}
	\right),~~~
	M^{(1)}=\left(
	\begin{array}{cc}
	0 & 0 \\
	d^\mu & 0 \\
	\end{array}
	\right), ~~~
	M^{(p)}=0,~~~(p\geq 2).
	\ee
For $\det M^{(0)}\ne0$, both $a$ and $c$ must not vanish.
In this case, \eqref{PDAE2} simplifies as
	\be
	a\e^1=0,~~~d^\mu\pa_\mu\e^1+b\e^1+c\e^2=0. \label{exmg2pdae}
	\ee
Now one finds $\e^1=0$ from the first equation.
After substituting it to the second equation, it is immediate to see $\e^2=0$.
Thus, we obtain a unique solution~$\e^I=0$ regardless of nonzero $M^{(1)}$.
Hence, the condition~\eqref{mg1cond} is too restrictive.

The mathematical criterion for the unique solvability of \eqref{PDAE2} is defined in the following manner.
The solution of \eqref{PDAE2} could be formally written as
\be \e^I = (\hat{Q}\cdot 0)^I, \label{Deltafsol} \ee
where $\hat{Q}^I_J$ is the inverse matrix operator of $\hat{P}^I_J$. 
In general, $\hat Q^I_J$ involves integral operators and thus the solution~\eqref{Deltafsol} contains integration constants.
However, if the PDAE system is uniquely solvable, there exists a derivative-operator-valued matrix $\hat{Q}^I_J$ satisfying
	\be
	\hat{P}^I_K\hat{Q}^K_J=\hat{Q}^I_K\hat{P}^K_J=\delta^I_J,
	\ee
and in this case~\eqref{Deltafsol} indeed gives a unique solution~$\e^I=0$.
The case of \eqref{mg1cond} with $m_g=1$, namely $\hat{P}=M^{(0)}$, is indeed the only case that the inverse operator~$\hat{Q}$ is independent of integral operators:~it is just given by $\hat{Q}=1/M^{(0)}$.
In general, there is no systematic way for judging explicitly the existence of such an inverse matrix~$\hat{Q}^I_J$ only involving derivative operators for any given $\hat{P}^I_J$.
Yet, there is a systematic way for some special cases.
We shall return to this point in \S \ref{sec:anm}.

The key feature of the unique solvability of a PDAE system is the fact that it is shared by its adjoint PDAE system, which is defined as follows.
Let $u_I$ and $v^I$ be arbitrary functions with compact support.
Then the adjoint of $\hat{P}^I_J$ is defined through integration by parts as
	\begin{align}
	\left\langle u_I,\hat{P}^I_Jv^J\right\rangle &=\int d^Dx\,u_I\sum_s M^{(s)}{}^I_J\pa_{(s)}v^J \nonumber \\
	&=\int d^Dx\left[\sum_s (-1)^s\pa_{(s)}\left(u_IM^{(s)}{}^I_J\right)\right]v^J\equiv\left\langle \hat{P}^\da{}^I_Ju_I,v^J\right\rangle, \label{PDAE3}
	\end{align}
where a dagger represents an adjoint operator.
Namely, $\hat{P}^\da{}^I_J$ acts on $u_I$ as
	\be
	\hat{P}^\da{}^I_Ju_I=\sum_s (-1)^s\pa_{(s)}\left(u_IM^{(s)}{}^I_J\right). \label{adPDAE}
	\ee
If $\hat Q^I_J$ involves integral operators, we cannot define $\hat Q^\da{}^I_J$ in the same way.
By contrast, if $\hat Q^I_J$ is written solely by differential operators, one can define the adjoint operator for $\hat{Q}^I_J$ in the same way as \eqref{PDAE3} and it can be easily shown that $\hat{Q}^\da{}^I_J$ is the inverse operator of $\hat{P}^\da{}^I_J$:
	\be
	\hat{P}^\da{}^I_K\hat{Q}^\da{}^K_J=\hat{Q}^\da{}^I_K\hat{P}^\da{}^K_J=\delta^I_J.
	\ee
Therefore, if \eqref{PDAE2} is uniquely solvable for $\e^J$, the adjoint PDAE system for functions $\lambda_I$,
	\be
	\hat{P}^\da{}^I_J\lambda_I=0,
	\ee
also has a unique solution, which is given by $\lambda_I=\hat{Q}^\da{}^J_I(\hat{P}^\da{}^K_J\lambda_K)=0$.

The above feature is precisely what we need to prove the Theorem.
If one imposes the gauge conditions~\eqref{gengfix} after deriving all the EOMs, one obtains
	\be \label{EOM1op} \E_i = 0 , \quad f^I=0, \ee
as basic equations to describe the dynamics of the system.
As emphasized before, the number of gauge conditions~$m_g$ is in general equal to or smaller than the total number of gauge symmetries~$m$.
On the other hand, if one fixes the gauge at the action level, one must minimize the action under $f^I=0$.
This can be achieved by the method of Lagrange multiplier.
Namely, we add to the Lagrangian the gauge-fixing functions~$f^I$ multiplied by $\lambda_I$:
	\be S_{\rm fix}=\int d^Dx ( L+\lambda_I f^I ). \label{gfaction} \ee
Here, $\lambda_I$ as well as $\phi^i$ are considered as dynamical variables.
Note that one may use the gauge conditions to modify the form of $L$, because such use of the gauge conditions is equivalent to redefinition of $\lambda_I$.
Yet, it is assumed for convenience that gauge conditions are not used in $L$ to eliminate some variables.
The EOMs derived from the gauge-fixed action~\eqref{gfaction} are given by
	\be \label{EOM2op} \E_i = - \sum_{q=0}^\ell (-1)^q \pa_{(q)} \mk{ \lambda_I u_i^{I(q)} } , \quad f^I=0, \ee
where $u_i^{I(q)}$ is defined in \eqref{eq-Ms}.
We can show that if a gauge fixing is complete in the sense that $\e^I=0$ is the unique solution for the PDAE system~\eqref{PDAE2}, then all the Lagrange multipliers~$\lambda_I$ are vanishing, i.e., \eqref{EOM2op} is equivalent to \eqref{EOM1op}.
Plugging \eqref{EOM2op} into the Noether identity~\eqref{fiden}, we note that 
	\be
	\sum_{p=0}^k \sum_{q=0}^\ell (-1)^{p+q} \pa_{(p)} \kk{ F^{i(p)}_J  \pa_{(q)} \mk{ \lambda_I u_i^{I(q)} }  }=0. \label{EOMfix}
	\ee
One can verify that the left-hand side is equivalent to $\hat{P}^\da{}^I_J \lambda_I$ defined by \eqref{adPDAE}.
Indeed, from the expression for $M^{(s)}{}^I_J$ given in \eqref{eq-Ms}, we have
	\begin{eqnarray}
	{\hat P}^\da{}^I_J\lambda_I&=&\sum_{s=0}^{k+\ell} {(-1)}^s \pa_{(s)} \left( \lambda_I M^{(s)}{}^I_J \right) \nonumber \\
	&=&\sum_{p=0}^k \sum_{q=0}^\ell \sum_{r=0}^q  \left(\begin{array}{c}
	q\\ r
	\end{array}\right)
	{(-1)}^{r+p} \pa_{(p)} \pa_{(r)} \left[ \lambda_I u_i^{I(q)} \pa_{(q-r)}
	F^{i(p)}_J \right] \nonumber \\
	&=& \sum_{p=0}^{k} \sum_{q=0}^\ell \sum_{r=0}^q \sum_{s=0}^r
	\left(\begin{array}{c}
	q\\ r
	\end{array}\right)
	\left(\begin{array}{c}
	r\\ s
	\end{array}\right)
	{(-1)}^{r+p} \pa_{(p)} \left[ \pa_{(s)} \left( \lambda_I u_i^{I(q)} \right) \pa_{(q-s)}
	F^{i(p)}_J \right] \nonumber \\
	&=&\sum_{p=0}^{k} \sum_{q=0}^\ell {(-1)}^{p+q}
	\pa_{(p)} \left[ F^{i(p)}_J \pa_{(q)} \left( \lambda_I u_i^{I(q)} \right) \right],
	\end{eqnarray}
where in going from the third line to the fourth line, we have interchanged the summations $\sum_{r=0}^q\sum_{s=0}^r=\sum_{s=0}^q\sum_{r=s}^q$ and used a formula
	\be
	\sum_{r=s}^q
	{(-1)}^r 
	\left(\begin{array}{c}
	q\\ r
	\end{array}\right)
	\left	(\begin{array}{c}
	r\\ s
	\end{array}\right) ={(-1)}^q \delta_{qs}.
	\ee
Therefore, \eqref{EOMfix} is the adjoint PDAE system to \eqref{PDAE2}, and as we mentioned earlier, they share the unique solvability.
Namely, if the gauge fixing is complete, there exists an inverse matrix of the adjoint operator~$\hat P^\da{}^I_J$, which leads to the unique solution~$\lambda_I=0$.
This completes the proof of the Theorem.

In conclusion, if the gauge fixing by the conditions~\eqref{gengfix} is complete, one could impose the gauge conditions at the action level and then derive EOMs without any inconsistency.
This is because the process yields the same set of EOMs obtained from varying the original action and then imposing the gauge conditions.
On the other hand, if the gauge fixing is incomplete and imposed at the action level, it may lead to an incorrect set of EOMs as some part of the EOMs are lost in general.
One could circumvent this situation by deriving EOMs from the original action without incomplete gauge fixing, and then impose the gauge conditions.
Another consistent way of analysis is that after deriving EOMs from an incompletely gauge-fixed action, one derives the lost EOMs from the original action without gauge fixing, and then impose the gauge conditions.
Then the combined set of EOMs is equivalent with the one obtained by imposing the gauge conditions after deriving EOMs.
In general, imposing the gauge conditions at the action level simplifies the derivation of EOMs since some part of the EOMs can be derived from a simplified action.
However, one needs to pay attention to supplement all the lost EOMs.

The remaining thing is how to check if the gauge fixing is complete, i.e., how to check the existence of the inverse matrix of $\hat P^I_J$. 
Although there is no systematic method for this in general, there are at least two exceptional cases:~one is $m_g=1$ and the other is $D=1$. 
The former is a field theory with a single gauge symmetry in $D$-dimensional spacetime, whereas the latter corresponds to analytical mechanics of a point particle in $n$-dimensional space.
In the case of $m_g=1$, as mentioned earlier, the existence of any derivative term spoils the unique solvability of the equation~$\hat{P}\e=0$.
Thus $\hat{P}$ must be of the form~\eqref{mg1cond}, $\hat{P}=M^{(0)}$.
We shall discuss the other case of $D=1$ in \S \ref{sec:anm}.

\subsection{Possible extension of the Theorem}
\label{ssec:bou}

As mentioned in the previous section, the Theorem postulates that (a part of) the gauge functions~$\xi^I$ are completely determined by the same number of conditions~$f^I=0$.
Therefore, it does not apply to, e.g., electrodynamics.
Indeed, the Maxwell theory in flat spacetime without source term
	\be
	L=-\f{1}{4}F_{\mu\nu}F^{\mu\nu}
	\ee
has a gauge symmetry under
	\be
	A_\mu(t,\vx)\to A_\mu(t,\vx)+\pa_\mu\xi(t,\vx),
	\ee
in which the gauge function~$\xi$ appears only with derivative.
This makes it impossible to determine $\xi$ uniquely for any gauge condition, i.e., any gauge fixing is incomplete, and thus the assumption of the Theorem could not be satisfied.

In cases where the gauge fixing is incomplete, one may consider imposing some additional conditions, which could be of the form~\eqref{gengfix} or possibly be boundary conditions, to fix the residual gauge DOFs.
Then, now that the gauge fixing is completed, the same result would hold:~the lost EOMs can be recovered.
Although this is not always the case (see \S \ref{ssec:toy}), one can indeed recover the lost EOMs in the case of the Coulomb gauge in electrodynamics.
Starting from a general configuration of $A^\mu$ that satisfies the boundary condition~$A^\mu\to 0$ as $|\vx|\to\infty$, one can always find $\xi$ so that the transformed variables fulfill the Coulomb gauge condition~$\pa_iA^i=0$:
	\be
	\Lap\!\xi(t,\vx)=-\pa_iA^i(t,\vx),
	\ee
where $\Lap\equiv\pa^i\pa_i$ is the Laplacian.
There still remains ambiguity of function that satisfies $\Lap\xi_h(t,\vx)=0$, and this DOF can be used to set $A^0=0$.
Since $A_0$ is transformed as $A_0\to A_0+\dot{\xi}_h$, we choose
	\be
	\xi_h(t,\vx)=\psi(\vx)-\int ^tdt'A_0(t',\vx),
	\ee
where $\psi(\vx)$ is an arbitrary harmonic function.
One can show that this $\xi_h$ satisfies $\Lap\xi_h=0$ by use of the EOM for $A^0$ with $\pa_iA^i=0$.
If we require that the boundary condition for $A^i$ is maintained by the gauge transformation, then one can fix $\psi(\vx)=0$ and the only remaining gauge DOF is a constant.

Now we consider fixing $\pa_iA^i=0$ and $A^0=0$ in the Lagrangian, namely,
	\be
	L_{\rm fix}=-\f{1}{4}F_{\mu\nu}F^{\mu\nu}+\lambda(\pa_iA^i)+\alpha A^0,
	\ee
and demonstrate the whole set of the EOMs for $L$ can be recovered.
The EOMs for the field~$A^\mu$ and the Lagrange multipliers~$\lambda,\alpha$ are given by
	\begin{align}
	\label{Coulomb1} \E_\mu&=\pa^\nu F_{\nu\mu}-\delta^i_\mu\pa_i\lambda+\delta^0_\mu\alpha=0, \\
	\label{Coulomb2} \E_\lambda&=\pa_iA^i=0, \\
	\label{Coulomb3} \E_\alpha&=A^0=0.
	\end{align}
The original set of EOMs~$\pa^\nu F_{\nu\mu}=0$ follows if $\lambda={\rm const.}$ and $\alpha=0$.
With \eqref{Coulomb2} and \eqref{Coulomb3}, one can simplify \eqref{Coulomb1} to get
	\be
	\E_0=\alpha=0,~~~\E_i=\Box A_i-\pa_i\lambda=0.
	\ee
Obviously, $\alpha=0$ from the first equation.
From the second equation and the boundary condition that $A^i\to 0$ as $|\vx|\to\infty$, one can conclude $\lambda\to {\rm const}$. 
On the other hand, from $\pa^i\E_i=0$ with \eqref{Coulomb2}, we obtain $\Lap\lambda=0$ and thus $\lambda={\rm const.}$ everywhere in spacetime.
Hence, we have recovered the original EOMs.

Similar arguments also hold in another choice of gauge fixing, e.g., Lorenz gauge condition $\partial_\mu A^\mu=0$ supplemented with an additional condition~$A^0=0$, and may be extended to some other gauge theories.
Nevertheless, we do not consider further generalization of the Theorem here.

\section{Criterion for complete gauge fixing}
\label{sec:anm}

In \S \ref{sec:fth}, we showed that gauge fixing at the action level is justified if the gauge fixing is complete. 
We saw that for the case of $m_g=1$, i.e., field theory with a single gauge symmetry, it is immediate to derive the necessary and sufficient condition~\eqref{mg1cond} for the gauge fixing to be complete.
In this section, we focus on the case of $D=1$, and show that it is possible to check if given gauge conditions define complete gauge fixing or not by transforming the corresponding ordinary DAEs (ODAEs) into some canonical form.

\subsection{Setup}

In $D=1$ case, the fields~$\phi^i$ are functions of time only and the system is equivalent to analytical mechanics of a point particle in $n$-dimensional space.
Therefore, we employ $q^i$ instead of $\phi^i$ as fundamental variables to emphasize this point and write the Lagrangian as $L=L(q^i, \dot q^i, \ddot q^i,\cdots;t)$ with $q^i=q^i(t)\,(i=1,\cdots, n)$.
Nevertheless, note that this class applies not only to analytical mechanics but also to field theories with a homogeneous configuration, field theories written in terms of Fourier decomposed variables, etc.  

As in \S \ref{sec:fth}, we assume that the Lagrangian is invariant up to total derivative under a gauge transformation
	\be q^i(t) \to q^i(t) + \Delta_\xi q^i ,  \ee
where $\Delta_\xi q^i$ involve higher-order time derivatives of the gauge functions~$\xi^I(t)\,(I=1,\cdots, m)$.
For an infinitesimal gauge transformation, the change of $q^i$ can be linearized as
	\be \label{hodgaugetrans} \Delta_\e q^i = \sum_{p=0}^kF^{(p)}{}^i_I(t)\f{d^p\e^I(t)}{dt^p}. \ee
The Noether identity~\eqref{fiden} for the EOMs~$\E_i$ then reads
	\be \label{anfiden} \sum_{p=0}^k (-1)^p \f{d^p}{dt^p} \left( \E_i F^{(p)i}_I \right) = 0. \ee
The gauge-fixing conditions depending on $q^i$ and their time derivatives take the form of
	\be	\label{ddgaugefix} f^I\left(q^i,\dot q^i,\ddot q^i,\cdots,\f{d^\ell q^i}{dt^\ell};t \right)=0, ~~~(I=1,\cdots, m_g). \ee
If the gauge fixing is complete, \eqref{ddgaugefix} fixes $m_g\,(\leq m)$ components of $\xi^I$ without ambiguity of integration constant.
Similarly to the analysis in \S \ref{sec:fth}, below we denote by $\xi^I$ only such $m_g$ components that are relevant to the gauge fixing.

\subsection{Derivation of the criterion}
\label{ssec:crit}

As we have proved in \S \ref{sec:mth}, there is a one-to-one correspondence between the completeness of gauge fixing and the unique solvability of \eqref{PDAE2}.
In our present case of $D=1$, \eqref{PDAE2} reads
	\be
	0=\Delta_\e f^I=M^{(0)}{}^I_J\e^J+\sum_{p=1}^{k+\ell}
	M^{(p)}{}^I_J\f{d^p\e^I(t)}{dt^p}, \label{hoDAE}
	\ee
where we have sorted terms by the order of derivative as in \eqref{psort}.
This is a system of linear ODAEs with higher-order derivatives.
Introducing auxiliary variables $\eta_{(p)}^I$ as
	\be
	\eta_{(1)}^I=\dot{\e}^I,~~~\eta_{(p)}^I=\dot{\eta}_{(p-1)}^I,~~~(p=2,3,\cdots,k),
	\ee
one has the following first-order DAE system for the set of variables $(\e^I, \eta_{(p)}^I)$:
	\be
	\begin{bmatrix}
	M^{(0)}{}^I_J&M^{(1)}{}^I_J&M^{(2)}{}^I_J&\cdots&M^{(k)}{}^I_J \\
	0&\delta^I_J&0&\cdots&0 \\
	0&0&\delta^I_J&\cdots&0 \\
	\vdots&\vdots&\vdots&\rotatebox{10}{$\ddots$}&\vdots \\
	0&0&0&\cdots&\delta^I_J
	\end{bmatrix}
	\begin{bmatrix}
	\e^J\\
	\eta_{(1)}^J\\
	\eta_{(2)}^J\\
	\vdots \\
	\eta_{(k+\ell)}^J
	\end{bmatrix}-
	\begin{bmatrix}
	0&0&0&\cdots&0 \\
	\delta^I_J&0&0&\cdots&0 \\
	0&\delta^I_J&0&\cdots&0 \\
	\vdots&\ddots&\ddots&\ddots&\vdots \\
	0&\cdots&0&\delta^I_J&0
	\end{bmatrix}
	\begin{bmatrix}
	\dot{\e}^J\\
	\dot{\eta}_{(1)}^J\\
	\dot{\eta}_{(2)}^J\\
	\vdots \\
	\dot{\eta}_{(k+\ell)}^J
	\end{bmatrix}=
	\begin{bmatrix}
	0\\
	0\\
	\vdots \\
	\vdots \\
	0
	\end{bmatrix}. \label{reducedDAE}
	\ee
In this way one can always reduce the higher-order DAEs~\eqref{hoDAE} for $\e^I$ to the manifestly first-order DAEs~\eqref{reducedDAE} for $(\e^I, \eta_{(p)}^I)$.
Note that \eqref{hoDAE} and \eqref{reducedDAE} share the unique solvability, since $\e^I$ and all the auxiliary variables $\eta_{(p)}^I$ vanish if and only if $\e^I=0$.
Therefore, without loss of generality, below we consider the unique solvability of the first-order DAE system of the form
	\be \label{eq-epsilon} \hat{P}^I_J\e^J\equiv M^I_J \dot \e^J + N^I_J \e^J=0 , \ee 
where $M$ and $N$ are time-dependent $m \times m$ matrices.

In the case of a single gauge symmetry, the criterion for the unique solvability was simple, i.e., all the coefficients of derivatives vanish.  
Therefore, one may think that $\epsilon^I$ are uniquely determined only when 
	\be M^I_J=0 \quad \text{and} \quad \det N^I_J\ne0. \label{suf1} \ee
However, as we mentioned earlier, this condition is too restrictive.
While \eqref{suf1} is a sufficient condition, it is not a necessary condition. 

Another uniquely solvable example is that
	\be \label{solform} M^I_J= K^I_J \quad \text{and} \quad N^I_J=\delta^I_J , \ee
where $K^I_J$ is a strictly lower (upper) triangular matrix, i.e., all the components of the matrix on and above (below) the diagonal are vanishing.
Note that $K^I_J$ can depend on time.
If $K^I_J$ is strictly lower triangular, one can first determine $\e^1$ uniquely.  
One then determines $\e^2$, as one can treat nonvanishing derivative $\dot \e^1$ as a source term. 
Likewise, one can continue to determine all the components of $\e^I$ uniquely.
On the other hand, if $K^I_J$ is strictly upper triangular, one can start from $\e^m$, and proceed to determine $\e^{m-1}, \e^{m-2}, \cdots, \e^1$ in order.
In terms of the operator matrix $\hat P$ in \eqref{eq-epsilon}, the case of \eqref{solform} amounts to
	\be \hat P ^I_J= \delta^I_J + K^I_J \f{d}{dt} . \ee
Clearly, $\hat{P}$ has an inverse matrix
	\be
	\Bigl(\hat{P}^{-1}\Bigr)^I_J=\delta^I_J+\sum_{s=1}^{m-1}\Bigl[\Bigl(-K\f{d}{dt}\Bigr)^s\,\Bigr]^I_J,
	\ee
as expected.
Indeed, as explained in \S \ref{sec:mth}, the existence of $\hat P^{-1}$ without integral is equivalent to the unique solvability of the corresponding system of equations.
Besides the case of \eqref{solform}, there are still other forms of $(M,N)$ for which \eqref{eq-epsilon} is uniquely solvable. 
For instance, $\delta^I_J$ can be relaxed to some diagonal matrix whose diagonal components are all nonvanishing, and then $N$ can be added by any strictly lower (upper) triangular matrix, which is a generalization of \eqref{exmg2}.
Therefore, the sufficient condition above can be generalized as
	\be \label{suf2} M^I_J = K^I_J \quad \text{and} \quad N^I_J = D^I_J + L^I_J, \ee
where $K^I_J$ and $L^I_J$ are time-dependent strictly lower (upper) triangular matrices, and $D^I_J$ is a time-dependent regular diagonal matrix.
In general, however, it is not possible to write down all the uniquely solvable cases in a simple form.

On the other hand, the following is necessary for the unique solvability (see Appendix~\ref{DAE}):
	\be
	\det M^I_J=0. \label{MN}
	\ee
If this condition is not satisfied, the system~\eqref{eq-epsilon} is basically a set of ordinary differential equations (ODEs) and cannot be solved uniquely for $\e^I$.
Although the above condition is necessary for the unique solvability, it is not a sufficient condition and does not necessarily guarantee the unique solvability. 

In practice, as we shall see in \S \ref{ssec:grp}, the sufficient conditions~\eqref{suf1} and \eqref{suf2}, and the necessary condition~\eqref{MN} are powerful enough when one considers specific gauge theories with multiple gauge symmetries and judges whether a gauge fixing is complete or not.
In addition, by using a transformation of a given ODAE system to some canonical form, it is possible to check the unique solvability of the ODAE system with a general pair of matrices $(M,N)$ (see Appendix~\ref{DAE}).

In summary, it is always possible to judge whether any given gauge fixing is complete or not in the case of $D=1$.
Practically, one can first check the sufficient conditions~\eqref{suf1} and \eqref{suf2}, and the necessary condition~\eqref{MN}.
If none of them are useful to determine whether the system is uniquely solvable, one can proceed to perform the methodology in Appendix~\ref{DAE}, which always works.
From the above argument, now it is clear why we restrict ourselves to analytical mechanics.
In the case of field theories with $D\ge 2$, one can still reduce any higher-order PDAE system to a first-order system as \eqref{eq-epsilon}:
	\be
	M^\mu{}^I_J\pa_\mu\e^J+N^I_J \e^J=0,
	\ee
which is characterized by the set of $D+1$ matrices $(M^\mu,N)$.
Even in this case, one could still consider sufficient conditions and a necessary condition similar to \eqref{suf1}, \eqref{suf2}, and \eqref{MN} (see \S \ref{ssec:ssspert}).
However, to the best of our knowledge, the criterion for the unique solvability of a general PDAE system is still an open issue.

\section{Applications to scalar-tensor theories}
\label{sec:app}

In \S \ref{sec:fth} and \S \ref{sec:anm}, we have established a general theorem that guarantees the validity of gauge fixing at the action level so long as the gauge fixing is complete.
In this section, we apply the obtained results to some dynamical systems in the framework of generic scalar-tensor theories for demonstration.
We shall also see that imposing incomplete gauge fixing at the action level leads the subsequent analysis to some inconsistency.

Throughout this section, we focus on a general scalar-tensor theory in $D$-dimensional spacetime
	\be
	S=\int d^Dx\sqrt{-g}L(g_{\mu\nu},\pa_\lambda g_{\mu\nu},\pa_\lambda\pa_\sigma g_{\mu\nu},\cdots;\phi,\pa_\lambda\phi,\pa_\lambda\pa_\sigma\phi,\cdots), \label{generalST}
	\ee
which possesses general covariance, i.e., the action is invariant under an infinitesimal transformation of coordinates~$x^\mu\to x^\mu+\e^\mu$.
The gauge transformation of the metric and the scalar field is then given by
	\be
	\begin{split} 
	g_{\mu\nu} &\to g_{\mu\nu} - \nabla_\mu \e_\nu - \nabla_\nu \e_\mu , \\
	\phi &\to \phi-\e^\mu\nabla_\mu \phi . \label{gencov}
	\end{split}
	\ee
Indeed, the gauge transformation of the Lagrangian density becomes total derivative:
	\be
	\Delta_\e(\sqrt{-g}L)=(-\sqrt{-g}\nabla_\mu\e^\mu)L+\sqrt{-g}(-\e^\mu\nabla_\mu L)=-\sqrt{-g}\nabla_\mu(\e^\mu L)=-\pa_\mu(\e^\mu\sqrt{-g}L).
	\ee

\subsection{Homogeneous and isotropic universe}
\label{ssec:grb}

Now we work on the flat Friedmann-Lema\^{i}tre-Robertson-Walker (FLRW) metric
	\be ds^2 = -N^2(t) dt^2 + a^2(t) \delta_{ij} dx^idx^j, \label{FLRW} \ee
with $\phi=\phi(t)$.\footnote{One can verify that imposing the metric ansatz~\eqref{FLRW} at the action level is harmless, i.e., it yields the same set of equations as the one obtained by imposing the ansatz after deriving EOMs.
This statement can be extended to general isometries \cite{Palais:1979rca}.}
Taking $\e^0=\e^0(t)$ and $\e^i=0$ in \eqref{gencov}, we have the following gauge transformation for $N,a$, and $\phi$:
	\begin{align} 
	\Delta_\e N&=-N\dot{\e}^0-\dot{N}\e^0, \notag \\
	\Delta_\e a&=-\dot{a}\e^0, \\
	\Delta_\e\phi&=-\dot{\phi}\e^0 .\notag
	\end{align}
With this transformation rule, one can write down the Noether identity~\eqref{anfiden} as
	\be
	N\dot{\E}_N-\dot{a}\E_a-\dot{\phi}\E_\phi=0, \label{grbgeomsum}
	\ee
where $\E_N,\E_a,\E_\phi$ are the EOMs for $N,a,\phi$, respectively.
Note that the above identity holds for any spacetime dimension~$D$.

As a simple case, let us consider general relativity with a canonical scalar field in four dimensions:
	\be \label{GRscalar} S_{\rm GR}=\int d^4x \sqrt{-g}\kk{  \f{\Mpl^2}{2}(R-2\Lambda) -\f{\omega}{2} g^{\mu\nu} \pa_\mu \phi \pa_\nu \phi - V(\phi)  } .  \ee
In the present case of the flat FLRW spacetime, the Lagrangian reads
	\be \label{GRbg} \sqrt{-g}L = a^3N \kk{  \Mpl^2 \mk{ 3\f{\ddot a}{aN^2} + 3 \f{\dot a^2}{a^2N^2} - 3 \f{\dot a}{aN} \f{\dot N}{N^2} -\Lambda  } + \f{\omega}{2}\f{\dot \phi^2}{N^2}-V(\phi)  }. \ee
The EOMs for $N,a,\phi$ are respectively given by
	\begin{align} \label{grbgeom}
	\E_N &=  a^3 \kk{  \Mpl^2\mk{ 3\f{\dot a^2}{a^2N^2} -\Lambda }  - \mk{\f{\omega}{2}\f{\dot \phi^2}{N^2} +V(\phi) } },  	\notag\\
	\E_a &= 3a^2 N \kk{  \Mpl^2  \mk{ 2\f{\ddot a}{aN^2} +\f{\dot a^2}{a^2N^2} - 2\f{\dot a }{aN} \f{\dot N}{N^2} - \Lambda} + \f{\omega}{2}\f{\dot \phi^2}{N^2} - V(\phi)  }, \\
	\E_\phi &= -a^3 N \kk{  \omega\mk{ \f{\ddot \phi}{N^2} + 3\f{\dot a}{aN}\f{\dot \phi}{N} - \f{\dot \phi}{N}\f{\dot N}{N^2}  } + V'(\phi)  },\notag
	\end{align}
and they indeed satisfy the identity~\eqref{grbgeomsum}.

If one gauge fixes either $a$ or $\phi$ at the action level, one would not obtain $\E_a$ or $\E_\phi$.
This does not cause any problem because the lost EOM $\E_a$ or $\E_\phi$ can be recovered from the other EOMs by using \eqref{grbgeomsum}, which means that either $\E_a$ or $\E_\phi$ is a redundant equation.
This result reflects the fact that fixing of either $a$ or $\phi$ is complete gauge fixing as $\Delta_\e a=0$ or $\Delta_\e\phi=0$ has the unique solution $\e^0=0$.

However, if one gauge fixes $N$ at the action level, one would lose $\E_N$ and cannot recover it from the other EOMs, which is clear as \eqref{grbgeomsum} involves $\dot \E_N$.
To avoid this situation, one should not fix $N$ at the action level.
One should fix $N$ only after deriving the EOM for $N$.
Although this is a widely used strategy for gauge fixing of $N$, to the best of our knowledge, its reason had not been sufficiently investigated and the general criterion had not been clarified.
Now it is clear that one can use complete gauge fixing at the action level without losing EOMs.
For incomplete gauge fixing, one should circumvent to use it until one derives EOMs.  

\subsection{Linear cosmological perturbations}
\label{ssec:grp}

We consider  the scalar perturbations around the FLRW metric (see e.g. \cite{Kodama:1985bj, Weinberg:2008})
	\be ds^2 = -N^2(1+2\Phi) dt^2 + 2 a N \pa_i B dt dx^i + a^2 \kk{ (1+2\Psi )\delta_{ij} + \mk{ \pa_i\pa_j -\f{\pa^2}{D-1} \delta_{ij}} E } dx^idx^j , \ee
and the perturbation of the scalar field $\delta \phi$.
From now on we work in the Fourier space.
Note that the perturbation variables with different wave numbers are decoupled.
The gauge transformation of the scalar perturbations corresponding to the coordinate redefinition by $x^\mu\to x^\mu+\e^\mu$ with $\e_i=\pa_i \e^S$ is given by
	\begin{align} \label{gautgrp}
	\Delta_\e \Phi &=-\dot{\e}^0-\f{\dot{N}}{N}\e^0 , \notag\\
	\Delta_\e \Psi &=-\f{\dot{a}}{a}\e^0+ \f{k^2}{3a^2}\e^S, \notag\\
	\Delta_\e B &= \f{1}{aN} \mk{ - \dot \e^S +N^2\e^0 + \f{2\dot a}{a} \e^S  } , \\
	\Delta_\e E &= - \f{2}{a^2} \e^S, \notag\\
	\Delta_\e \delta\phi &=-\dot{\phi}\e^0 .\notag
	\end{align}
Therefore, the Noether identity~\eqref{anfiden} reads
	\be
	\begin{split}
	&\dot{\E}_\Phi-\f{\dot{N}}{N}\E_\Phi-\f{\dot{a}}{a}\E_\Psi+\f{N}{a}\E_B-\dot{\phi}\E_{\delta\phi}=0, \\
	&\f{d}{dt} \mk{\f{1}{aN}\E_B} + \f{k^2}{3a^2}\E_\Psi + \f{2\dot a}{a^2N}\E_B  - \f{2}{a^2}\E_E = 0.
	\end{split} \label{EOMsumsGRpert}
	\ee
These identities hold in any spacetime dimension~$D$.

In the simple case of general relativity plus a canonical scalar field~\eqref{GRscalar} in four dimensions, the Euler-Lagrange equations for $\Phi,\Psi,B,E$ and $\delta\phi$ are respectively given by
	\begin{align}
	\E_\Phi &= a^3N \left\{ \Mpl^2 \left[ (\Lambda-9H^2)\Phi+3(3H^2-\Lambda)\Psi+6H\f{\dot \Psi}{N} 
	+ \f{k^2}{a^2}\mk{2aHB+2\Psi+\f{1}{3}k^2 E } \right]\right. \notag\\
	&~~~~~~+ \left. \omega \f{\dot \phi}{N} \left[\f{3}{2}\f{\dot \phi}{N} (\Phi-\Psi)-\f{\dot{\delta\phi} }{N} \right] + V (\Phi-3\Psi) -V' \delta\phi \right\}  ,  \notag\\ 
	\E_\Psi &= 3a^3N \left\{ \Mpl^2 \left[ 2\f{\ddot \Psi}{N^2} +2H\mk{3\f{\dot \Psi}{N}-\f{\dot \Phi}{N}} -2\f{\dot N}{N^2}\f{\dot \Psi}{N} + \mk{ 2\f{\ddot a}{aN^2} + H^2 -2H\f{\dot N}{N^2} }(\Psi-\Phi) -\Lambda(\Psi+\Phi) \right. \right. \notag\\
	&~~~~~~+\left. \left. \f{2k^2}{3 a^2} \mk{\Phi+\Psi+2aHB+\f{a\dot B}{N}} +\f{k^4}{9a^2}E \right]
	+ \omega \f{\dot \phi}{N} \left[ \f{\dot{\delta\phi}}{N} + \f{\dot \phi}{2N} (\Psi-\Phi) \right] -V(\Phi+\Psi) -V'\delta\phi \right\}  ,  \notag\\
	\E_B &= k^2a^2N \left\{  \Mpl^2\left[ (3H^2-\Lambda)aB -2\f{\dot\Psi}{N} +2H\Phi-\f{k^2}{3}\f{\dot E}{N}  \right]
	- \omega \f{\dot\phi}{N} \left[ \delta\phi+\f{\dot\phi}{2N}aB \right] -VaB  \right\} , \\
	\E_E &= \f{k^4a^3N}{6} \left\{ \Mpl^2 \left[  \f{2}{a^2}(\Phi+\Psi) + \f{2}{a}\mk{\f{\dot B}{N} +2HB} -\f{\ddot E}{N^2} +\mk{ \f{\dot N}{N^2} -3H }\f{\dot E}{N}  \right.  \right. \notag\\
	&~~~~~~+ \left.\left.  2E\mk{-2\f{\ddot a}{aN^2} -H^2 +\Lambda+2H\f{\dot N}{N^2} } + \f{k^2}{3a^2}E \right]
	+ E \mk{-\omega\f{\dot\phi^2}{N^2} +2V}  \right\} , \notag\\
	\E_{\delta\phi} &= -a^3N \left\{ V''\delta\phi + V'(\Phi+3\Psi) + \omega \left[ \f{\ddot{\delta\phi}}{N^2} + \f{\dot{\delta\phi}}{N} \mk{3H-\f{\dot N}{N^2} } + \mk{ -\f{\ddot \phi}{N^2} - 3H\f{\dot \phi}{N} +\f{\dot \phi}{N}\f{\dot N}{N^2} }(\Phi-3\Psi) \right. \right. \notag\\
	&~~~~~~- \left.\left.  \f{\dot \phi}{N} \mk{ \f{\dot \Phi}{N}-3\f{\dot \Psi}{N} } + \f{k^2}{a^2} \mk{ \delta\phi + \f{\dot\phi}{N}aB }  \right] \right\} ,\notag
	\end{align}
where $H\equiv \dot a/(aN)$.
One can confirm that the identity~\eqref{EOMsumsGRpert} is satisfied for these EOMs, after using the background EOMs~\eqref{grbgeom}.

As we showed in \S \ref{sec:fth}, the recoverability of the lost EOMs and the completeness of the gauge fixing are equivalent as they are related through adjoint DAE systems.
Below we consider three gauge-fixing conditions commonly used in cosmology, whose (in)completeness can be checked by the sufficient conditions~\eqref{suf1} and \eqref{suf2}, and the necessary condition~\eqref{MN}.

\begin{itemize}

\item Comoving gauge: $E=0$, $\delta \phi=0$ (complete).

In this case, the gauge conditions are $f^I=(E,\delta\phi)$.  
To check if it is complete or not, we consider an infinitesimal gauge transformation from the comoving gauge, and impose $\Delta_\e f^I=(\Delta_\e E,\Delta_\e \delta\phi) = (0,0)$.  From \eqref{gautgrp} we obtain
	\be
	\bem 
	0 & - \f{2}{a^2} \\
	-\dot{\phi} & 0
	\eem
	\bem \e^0 \\ \e^S \eem 
	=
	\bem 0 \\0 \eem .
	\ee
As it does not involve derivatives of $\e^0$ or $\e^S$, it is obvious that $\e^0=\e^S=0$ is the unique solution, i.e., the gauge fixing is complete.
This is the case of \eqref{suf1}.
Consequently, EOMs for $E$ and $\delta\phi$ can be recovered from 
	\be
	\bem   
	0 & -\dot{\phi} \\
	-\f{2}{a^2} & 0 \\
	\eem
	\bem \E_E \\ \E_{\delta\phi} \eem
	=
	\text{(sum of the other EOMs)} .
	\ee

\item Newtonian gauge: $B=0$, $E=0$ (complete).\footnote{Here, we do not consider $k=0$ modes.}

Since $f^I=(B,E)$, we impose $\Delta_\e B=0$ and $\Delta_\e E = 0$ in \eqref{gautgrp}, and obtain
	\be
	\bem
	0 &	-\f{1}{aN} \\ 
	0 & 0
	\eem
	\bem \dot \e^0 \\ \dot \e^S \eem 
	+
	\bem
	\f{N}{a} & \f{2\dot a}{a^2N} \\ 
	0 & -\f{2}{a^2}
	\eem
	\bem \e^0 \\ \e^S \eem 
	=
	\bem 0 \\ 0 \eem ,
	\ee
which satisfies the sufficient condition~\eqref{suf2},
and $\e^0=\e^S=0$ is the unique solution.
The identity~\eqref{EOMsumsGRpert} reads
	\be
	\bem
	0 & 0 \\ 
	\f{1}{aN} & 0
	\eem 
	\bem \dot \E_B \\ \dot \E_E \eem
	+
	\bem \f{N}{a} & 0 \\ 
	\f{\dot a}{a^2N}-\f{\dot N}{aN^2} & -\f{2}{a^2}
	\eem
	\bem \E_B \\ \E_E \eem
	=
	\text{(sum of the other EOMs)}.
	\ee
As expected, it satisfies the same sufficient condition~\eqref{suf2}, and we can recover $\E_B$, and then $\E_E$.

\item Synchronous gauge: $\Phi=0$, $B=0$ (incomplete).

Setting $\Delta_\e \Phi=0$ and $\Delta_\e B = 0$ in \eqref{gautgrp}, we have
	\be
	\bem 
	-1 & 0 \\
	0 & -\f{1}{aN}
	\eem
	\bem \dot \e^0 \\ \dot \e^S \eem 
	+
	\bem
	-\f{\dot N}{N} & 0 \\ 
	\f{N}{a} & \f{2\dot a}{a^2N}
	\eem
	\bem \e^0 \\ \e^S \eem 
	=
	\bem 0 \\ 0 \eem .
	\ee
Clearly, it violates the necessary condition~\eqref{MN}, and the gauge fixing is incomplete.  From \eqref{EOMsumsGRpert}, one can also check that $\E_\Phi$ and $\E_B$ cannot be recovered from the other EOMs as the adjoint DAE system violates the same necessary condition~\eqref{MN}. 

\end{itemize}

To reiterate, so long as the gauge fixing is complete, one can fix the gauge at the action level without losing EOMs.
In addition, one can also partially fix the gauge completely, which amounts to the case $m_g<m$ mentioned in \S \ref{sec:mth}.
For example, one can fix $\delta \phi=0$ at the action level, which determines $\e^0$ completely, and then derive the set of EOMs.
It is clear that one can recover $\E_{\delta\phi}$ from the other EOMs and there is no lost independent EOM.

\subsection{Spherically symmetric spacetime}
\label{ssec:sss}

Let us consider time-dependent spherically symmetric spacetime 
	\be
	ds^2=-A(t,r)dt^2+\f{dr^2}{B(t,r)}+2C(t,r)dtdr+E(t,r)r^2\gamma_{ij}dx^idx^j, \label{spht}
	\ee
with $\phi=\phi(t,r)$.\footnote{Similarly in \S \ref{ssec:grb}, one is allowed to impose the metric ansatz~\eqref{spht} at the action level.}
Here, $i,j$ label angular variables, and $\gamma_{ij}$ represents the metric of a $(D-2)$-dimensional maximally symmetric space with spatial curvature $\kappa=1$.
Let us perform the coordinate transformation $x^\mu\to x^\mu+\e^\mu$ with $\e^0=\e^0(t,r),\e^r=\e^r(t,r)$ and $\e^i=0$ for angular parts.
The gauge transformation of $A,B,C,E$ and $\phi$ is defined as
	\begin{align}
	\Delta_\e A&= -2A\dot\e^0 + 2C\dot\e^r -\dot A\e^0 -A'\e^r, \nonumber \\
	\Delta_\e B&= 2B^2C\e^0{}'+2B\e^r{}'-\dot B\e^0-B'\e^r, \nonumber \\
	\Delta_\e C&= -C\dot\e^0 -\f{1}{B}\dot\e^r + A\e^0{}'-C\e^r{}'- \dot C \e^0 - C'\e^r, \label{gtssst} \\
	\Delta_\e E&= -\dot E \e^0 - \f{(Er^2)'}{r^2} \e^r , \nonumber \\
	\Delta_\e \phi&= -\dot\phi\e^0 -\phi'\e^r. \nonumber
	\end{align}
It is commonly used to impose a metric ansatz by $C(t,r)=0$ and $E(t,r)=1$ (or sometimes $E(t,r)=1/B(t,r)$) in the action.  
However, this ansatz is not a complete gauge fixing.\footnote{Note that $E(t,r)=1$ alone is a complete gauge fixing. To see this, we set $\e^0=0$ in \eqref{gtssst} since $E(t,r)=1$ is achieved by redefinition of $r$ only. Then, $\Delta_\e E=0$ has the unique solution $\e^r=0$. Likewise one can also fix $\phi$ by complete gauge fixing.  Therefore, one can fix $E$ and/or $\phi$ by complete gauge fixing.}
Following the general prescription, if one sets $\Delta_\e C=\Delta_\e E=0$ in \eqref{gtssst}, it is clear that one cannot determine the set $(\e^0,\e^r)$ uniquely.  
Therefore, if one imposes the metric ansatz \eqref{spht} with $C(t,r)=0$ and $E(t,r)=1$ at the action level, the subsequent analysis does not capture the correct number of DOFs in general.  We shall return to this point in \S \ref{comments}.  

On the other hand, the static spherically symmetric spacetime
	\be
	ds^2=-A(r)dt^2+\f{dr^2}{B(r)}+2C(r)dtdr+E(r)r^2\gamma_{ij}dx^idx^j, \label{sph}
	\ee
with $\phi=\phi(r)$ allows us to fix $C(r)=0$ and $E(r)=1$ by complete gauge fixing.
For the coordinate transformation~$x^\mu\to x^\mu+\e^\mu$ with $\e^0=\e^0(r),\e^r=\e^r(r)$, and $\e^i=0$, the gauge transformation~\eqref{gtssst} simplifies as
	\begin{align}
	\Delta_\e A&=-A'\e^r, \nonumber \\
	\Delta_\e B&=2B^2C\e^0{}'+2B\e^r{}'-B'\e^r, \nonumber \\
	\Delta_\e C&=A\e^0{}'-C\e^r{}'-C'\e^r, \label{gtsss} \\
	\Delta_\e E&=-\f{(Er^2)'}{r^2}\e^r, \nonumber \\
	\Delta_\e \phi&=-\phi'\e^r. \nonumber
	\end{align}
The crucial difference from \eqref{gtssst} is that $\e^0$ appears only with radial derivative in \eqref{gtsss}, which is the consequence of the static ansatz of the spacetime.
As we mentioned in \S \ref{sec:mth}, such a choice of the gauge function is inappropriate.
It is not $\e^0$ itself but $\eta\equiv\e^0{}'$ that should be treated as a generator of the gauge transformation.
Then, the Noether identity~\eqref{anfiden} gives the following relations between the EOMs:
	\be
	\begin{split}
	&2B^2C\E_B+A\E_C=0, \\
	&2B\E_B'-C\E_C'+A'\E_A+3B'\E_B+\f{(Er^2)'}{r^2}\E_E+\phi'\E_\phi=0.
	\end{split} \label{sssnoet}
	\ee
Let us consider fixing $C(r)=0$ and $E(r)=1$ in the action.
Correspondingly, we set $\Delta_\e C=\Delta_\e E=0$ in \eqref{gtsss} and obtain
	\be
	\bem
	A & 0 \\
	0 & \f{2}{r}
	\eem
	\bem \eta \\ \e^r \eem=
	\bem 0 \\ 0 \eem,
	\ee
which has the unique solution $(\eta,\e^r)=(0,0)$.
Therefore, this gauge fixing is complete and thus $\E_C$ and $\E_E$ can be recovered from the Noether identity.\footnote{Of course, in addition to the gauge fixing of $C$ and $E$, \eqref{gtsss} tells us that one could consider other kind of complete gauge fixing.  In general one can fix any of $\{ A, E, \phi \}$ and/or either of $\{ B, C \}$ of \eqref{sph} by complete gauge fixing.}
However, the circumstance is a little different from the former examples.
In this case, setting $C=0$ in the first equation of \eqref{sssnoet} yields $\E_C=0$, which means $\E_C$ vanishes identically.
After that, $\E_E$ can be written in terms of the other EOMs by use of the second equation.

In the rest of this section, we investigate spherically symmetric solutions starting from a gauge-fixed action to illustrate the (in)appropriateness of (in)complete gauge fixing at the action level.
Here we consider the Einstein-Hilbert action with a cosmological constant in $D$ dimensions:
	\be
	S_{\rm EH}=\f{\Mpl^{D-2}}{2}\int d^Dx\sqrt{-g}(R-2\Lambda). \label{einhil}
	\ee
If we substitute the metric ansatz~\eqref{spht} with $C(t,r)=0$ and $E(t,r)=1$, the action becomes of the form
	\be
	S_{\rm EH}\bigr|_{C,E}=\f{D-2}{\Gamma\left(\f{D-1}{2}\right)}\pi^{\f{D-1}{2}}\Mpl^{D-2}\int dtdr\sqrt{\f{A}{B}}\left[\Bigl(r^{D-3}(1-B)\Bigr)'-\f{2\Lambda}{D-2}r^{D-2}\right].
	\ee
Hence we obtain EOMs as
	\be
	\Bigl(r^{D-3}(1-B)\Bigr)'-\f{2\Lambda}{D-2}r^{D-2}=0,~~~\left(\f{A}{B}\right)'=0, \label{spheom1}
	\ee
which yield the following solution:
	\be
	B(t,r)=1-\f{c_1(t)}{r^{D-3}}-\f{2\Lambda}{(D-1)(D-2)}r^2,~~~A(t,r)=c_2(t)B(t,r). \label{sphsol1}
	\ee
This result is obviously incompatible with cosmological-constant case of Birkhoff's theorem~\cite{Eiesland:1925}, according to which the coefficients~$c_1$ and $c_2$ must be constant.
This contradiction precisely originates from imposing the incomplete gauge-fixing condition~$C(t,r)=0$ and $E(t,r)=1$ at the action level.
Regarding this point, \cite{Deser:2003up} derived the corresponding set of EOMs in a similar manner, but finally neglected the time dependence of the solution~\eqref{sphsol1}.
Their solution is actually a physically correct one, but the above argument does not allow one to drop the time dependence of the solution.

A correct solution is obtained in the following manner.
Let us impose the metric ansatz~\eqref{spht} with $E(t,r)=1$ in the action~\eqref{einhil}, while $C(t,r)=0$ is imposed after deriving the three EOMs corresponding to the metric functions~$A,B$ and $C$.
Thus we start from the following action:
	\be
	S_{\rm EH}\bigr|_E
	=\f{D-2}{\Gamma\left(\f{D-1}{2}\right)}\pi^{\f{D-1}{2}}\Mpl^{D-2}\int dtdr\sqrt{\f{F}{B}}\left\{\Bigl(r^{D-3}(1-B)\Bigr)'-r^{D-3}\left[\f{2\Lambda}{D-2}r+\f{C}{F}\dot{B}+\f{B^3C^2}{2F^2}\left(\f{F}{B}\right)'\right]\right\}, \label{einhilfix}
	\ee
where $F\equiv A+BC^2$.
Note that one can fix $E(t,r)=1$ by redefinition of $r$, which can be read off from \eqref{gtssst}.
As we mentioned earlier, $E(t,r)=1$ alone is a complete gauge fixing and thus there is no residual DOF for redefining $r$.
On the other hand, if one fixes $C(t,r)=0$ by redefinition of $t$, then one can still redefine $t$ by some function that only depends on $t$:~$t\to \ti{t}(t)$.
The EOMs derived from \eqref{einhilfix} are
	\be
	\Bigl(r^{D-3}(1-B)\Bigr)'-\f{2\Lambda}{D-2}r^{D-2}=0,~~~\left(\f{A}{B}\right)'=0,~~~\dot{B}=0, \label{spheom2}
	\ee
where we have substituted $C=0$.
Since the first two equations coincide with \eqref{spheom1}, one obtains the same solution as in \eqref{sphsol1} from them.
The third equation is the difference from \eqref{spheom1}, which yields $c_1={\rm const}$.
As for $c_2(t)$, we can use the residual gauge DOF, namely, $t\to\ti{t}(t)$:~one can fix $c_2=1$ by choosing the new time coordinate~$\ti{t}$ so that $d\ti{t}=c_2(t)^{1/2}dt$.
Thus we obtain a solution which is consistent with Birkhoff's theorem.

Another consistent way of analysis is to derive EOMs from an incompletely gauge-fixed action, and to derive lost EOMs from the action without incomplete gauge fixing.
Then the combined set of EOMs yields a consistent analysis.
In the present example, after obtaining \eqref{spheom1}, one could derive the lost EOM for $C$ from \eqref{einhilfix}, and then impose $C=0$.
The resultant set of EOMs is the same as \eqref{spheom2}.

\subsection{Perturbations around static spherically symmetric background}
\label{ssec:ssspert}

Let us consider perturbations around the static spherically symmetric metric~\eqref{sph}.
In the following argument, we set $C(r)=0$ and $E(r)=1$ from the beginning.
We start from a brief review of the formalism to decompose the metric perturbations in general spacetime dimension developed in \cite{Kodama:2000fa}.
Any metric perturbation $h_{\mu\nu}\equiv g_{\mu\nu}-g^{(0)}_{\mu\nu}$ can be decomposed as follows:
	\begin{align}
	h_{ab}&=f_{ab}^{({\rm S})}(t,r)\bS, \nonumber \\
	h_{ai}&=r\left(f_a^{({\rm S})}(t,r)\bS_i+f_a^{({\rm V})}(t,r)\bV_i\right), \\
	h_{ij}&=2r^2\left(H_L^{({\rm S})}(t,r)\gamma_{ij}\bS+H_T^{({\rm S})}(t,r)\bS_{ij}+H_T^{({\rm V})}(t,r)\bV_{ij}+H_T^{({\rm T})}(t,r)\bT_{ij}\right), \nonumber
	\end{align}
where $a,b=(t,r)$ and $i,j$ denote angular variables.
On the other hand, a perturbation of the scalar field is written as
	\be
	\delta \phi=\delta \phi^{({\rm S})}(t,r)\bS.
	\ee
For the definitions of the harmonic functions~$\bS,\bS_i,\bS_{ij},\bV_i,\bV_{ij}$ and $\bT_{ij}$, see Appendix~\ref{hf}.
The expansion coefficients~$f_{ab}^{({\rm S})},f_a^{({\rm S})},H_L^{({\rm S})},H_T^{({\rm S})},f_a^{({\rm V})},H_T^{({\rm V})},H_T^{({\rm T})}$ represent the dynamical DOFs of the perturbation, and the superscripts denote the transformation property under rotations in the $(D-2)$-dimensional space:~$({\rm S}),({\rm V}),({\rm T})$ denote scalar, vector, tensor, respectively.\footnote{In four dimensions, the scalar perturbations are often referred to as even or $E$ modes, the vector perturbations are odd or $B$ modes, and the tensor perturbation is absent as the tensor harmonics $\bT_{ij}$ vanish.}
These three types of perturbations are completely decoupled.
Note that the coefficients~$f_a^{({\rm S})},f_a^{({\rm V})}$ of the harmonic vectors can be defined only for $\ell\ge1$, and the coefficients~$H_T^{({\rm S})},H_T^{({\rm V})},H_T^{({\rm T})}$ of the harmonic tensors appears only if $\ell\ge2$ (see Appendix~\ref{hf}).
Note also that we omitted the multipole index for the harmonic functions as each mode evolves independently.

The infinitesimal change $\e^\mu$ of coordinates can also be decomposed by use of the harmonic functions as
	\be
	\e_a=T_a^{({\rm S})}\bS,~~~\e_i=r\left(L^{({\rm S})}\bS_i+L^{({\rm V})}\bV_i\right). \label{gaugedec}
	\ee
With these functions $T_a^{({\rm S})},L^{({\rm S})}$ and $L^{({\rm V})}$, we obtain the gauge transformation of the dynamical variables.
For the scalar perturbations, we find
	\be
	\begin{split}
	\Delta_\e f_{tt}^{({\rm S})}&=-2\dot{T}_t^{({\rm S})}+A'BT_r^{({\rm S})}, \\
	\Delta_\e f_{tr}^{({\rm S})}&=-\dot{T}_r^{({\rm S})}-T_t'+\f{A'}{A}T_t^{({\rm S})}, \\
	\Delta_\e f_{rr}^{({\rm S})}&=-2T_r'^{({\rm S})}-\f{B'}{B}T_r^{({\rm S})}, \\
	\Delta_\e f_t^{({\rm S})}&=-\dot{L}^{({\rm S})}+\f{k_{\rm S}}{r}T_t^{({\rm S})}, \\
	\Delta_\e f_r^{({\rm S})}&=-L^{({\rm S})}{}'+\f{L^{({\rm S})}}{r}+\f{k_{\rm S}}{r}T_r^{({\rm S})}, \\
	\Delta_\e H_L^{({\rm S})}&=-\f{k_{\rm S}}{(D-2)r}L^{({\rm S})}-\f{B}{r}T_r^{({\rm S})}, \\
	\Delta_\e H_T^{({\rm S})}&=\f{k_{\rm S}}{r}L^{({\rm S})}, \\
	\Delta_\e \delta\phi^{({\rm S})}&=-B\phi'T_r^{({\rm S})},
	\end{split} \label{gtsca}
	\ee
where $k_{\rm S}^2$ is the eigenvalue of the scalar harmonic function $\bS$ (see Appendix~\ref{hf}).
For the vector perturbations,
	\be
	\begin{split}
	\Delta_\e f_t^{({\rm V})}&=-\dot{L}^{({\rm V})}, \\
	\Delta_\e f_r^{({\rm V})}&=-L^{({\rm V})}{}'+\f{L^{({\rm V})}}{r}, \\
	\Delta_\e H_T^{({\rm V})}&=\f{k_{\rm V}}{r}L^{({\rm V})}.
	\end{split} \label{gtvec}
	\ee
Here, $k_{\rm V}^2$ is the eigenvalue of the harmonic vector $\bV_i$.
The tensor perturbation $H_T^{({\rm T})}$ is invariant under the transformation~\eqref{gaugedec}:
	\be
	\Delta_\e H_T^{({\rm T})}=0.
	\ee
Now we are ready to write down the Noether identity.
Since the dynamical variables are functions of $(t,r)$, we employ the expression~\eqref{fiden} for multidimensional field theory.
For the scalar-type gauge functions $T_a^{({\rm S})}$ and $L^{({\rm S})}$, we obtain
	\be
	\begin{split}
	&2\dot{\E}_{tt}^{({\rm S})}+\E_{tr}^{({\rm S})}{}'+\f{A'}{A}\E_{tr}^{({\rm S})}+\f{k_{\rm S}}{r}\E_t^{({\rm S})}=0, \\
	&\dot{\E}_{tr}^{({\rm S})}+2\E_{rr}^{({\rm S})}{}'+A'B\E_{tt}^{({\rm S})}-\f{B'}{B}\E_{rr}^{({\rm S})}+\f{k_{\rm S}}{r}\E_r^{({\rm S})}-\f{B}{r}\E_L^{({\rm S})}-B\phi'\E_{\delta\phi}^{({\rm S})}=0, \\
	&\dot{\E}_t^{({\rm S})}+\E_r^{({\rm S})}{}'+\f{1}{r}\E_r^{({\rm S})}-\f{k_{\rm S}}{(D-2)r}\E_L^{({\rm S})}+\f{k_{\rm S}}{r}\E_T^{({\rm S})}=0,
	\end{split} \label{noetsca}
	\ee
and for the vector-type gauge function $L^{({\rm V})}$,
	\be
	\dot{\E}_t^{({\rm V})}+\E_r^{({\rm V})}{}'+\f{1}{r}\E_r^{({\rm V})}+\f{k_{\rm V}}{r}\E_T^{({\rm V})}=0. \label{noetvec}
	\ee
Here $\E_{\delta\phi}^{({\rm S})}$ is the EOM for $\delta\phi^{({\rm S})}$, and otherwise $\E_X^{(Y)}$ denotes the EOM for the expansion coefficient of the metric perturbation with the same indices.

In what follows, we consider three sets of (partial) complete gauge-fixing conditions and demonstrate the Theorem indeed holds.
The first two sets correspond to Regge-Wheeler gauge~\cite{Regge:1957td}, which is commonly used in the context of black-hole perturbation theory in four dimensions.

\begin{itemize}
\item $H_T^{({\rm V})}=0$:\\
For $H_T^{({\rm V})}$ to be defined appropriately, we focus on modes with $\ell\ge2$.
This gauge fixing is complete since $\Delta_\e H_T^{({\rm V})}=0$ in \eqref{gtvec} has the unique solution $L^{({\rm V})}=0$ since $k_{\rm V}\ne 0$.
As a result, the corresponding EOM $\E_T^{({\rm V})}$ can be recovered from the other EOMs by virtue of the Noether identity~\eqref{noetvec}.

\item $f_t^{({\rm S})}=H_L^{({\rm S})}=H_T^{({\rm S})}=0$: \\
Here, we restrict ourselves to modes with $\ell\ge2$ so that one can define both $f_t^{({\rm S})}$ and $H_T^{({\rm S})}$.
To check the completeness of the gauge fixing, we set $\Delta_\e f_t^{({\rm S})}=\Delta_\e H_L^{({\rm S})}=\Delta_\e H_T^{({\rm S})}=0$ in \eqref{gtsca}:
	\be
	\bem
	0 & 0 & -1 \\
	0 & 0 & 0 \\
	0 & 0 & 0
	\eem
	\bem \dot{T}_t^{({\rm S})} \\ \dot{T}_r^{({\rm S})} \\ \dot{L}^{({\rm S})} \eem+
	\bem
	\f{k_{\rm S}}{r} & 0 & 0 \\
	0 & -\f{B}{r} & -\f{k_{\rm S}}{(D-2)r} \\
	0 & 0 & \f{k_{\rm S}}{r}
	\eem
	\bem T_t^{({\rm S})} \\ T_r^{({\rm S})} \\ L^{({\rm S})} \eem=
	\bem 0 \\ 0 \\ 0 \eem.
	\ee
This satisfies the sufficient condition~\eqref{suf2} since $k_S\ne 0$.
Therefore, this gauge fixing is complete and one can recover $\E_t^{({\rm S})},\E_L^{({\rm S})}$ and $\E_T^{({\rm S})}$ from the Noether identity~\eqref{noetsca}:
	\be
	\bem
	0 & 0 & 0 \\
	0 & 0 & 0 \\
	1 & 0 & 0
	\eem
	\bem \dot{\E}_t^{({\rm S})} \\ \dot{\E}_L^{({\rm S})} \\ \dot{\E}_T^{({\rm S})}\eem+
	\bem
	\f{k_{\rm S}}{r} & 0 & 0 \\
	0 & -\f{B}{r} & 0 \\
	0 & -\f{k_{\rm S}}{(D-2)r} & \f{k_{\rm S}}{r}
	\eem
	\bem \E_t^{({\rm S})} \\ \E_L^{({\rm S})} \\ \E_T^{({\rm S})} \eem=\text{(sum of the other EOMs)}.
	\ee

\item $f_t^{({\rm S})}=f_r^{({\rm S})}=H_T^{({\rm S})}=0$: \\
Here again we consider modes with $\ell\ge2$.
We impose $\Delta_\e f_t^{({\rm S})}=\Delta_\e f_r^{({\rm S})}=\Delta_\e H_T^{({\rm S})}=0$ in \eqref{gtsca} to obtain
	\be
	\bem
	0 & 0 & -1 \\
	0 & 0 & 0 \\
	0 & 0 & 0
	\eem
	\bem \dot{T}_t^{({\rm S})} \\ \dot{T}_r^{({\rm S})} \\ \dot{L}^{({\rm S})} \eem+
	\bem
	0 & 0 & 0 \\
	0 & 0 & -1 \\
	0 & 0 & 0
	\eem
	\bem T_t^{({\rm S})}{}' \\ T_r^{({\rm S})}{}' \\ L^{({\rm S})}{}' \eem+
	\bem
	\f{k_{\rm S}}{r} & 0 & 0 \\
	0 & \f{k_{\rm S}}{r} & \f{1}{r} \\
	0 & 0 & \f{k_{\rm S}}{r}
	\eem
	\bem T_t^{({\rm S})} \\ T_r^{({\rm S})} \\ L^{({\rm S})} \eem=
	\bem 0 \\ 0 \\ 0 \eem.
	\ee
This example serves an application of the generalization of the sufficient condition~\eqref{suf2} to higher dimensions which we mentioned at the end of \S \ref{ssec:crit}.
Since $k_{\rm S}\ne 0$, this system has a unique solution and thus the gauge fixing is complete.
Focusing on the relevant components of the EOMs, the Noether identity~\eqref{noetsca} is written as
	\be
	\bem
	0 & 0 & 0 \\
	0 & 0 & 0 \\
	1 & 0 & 0
	\eem
	\bem \dot{\E}_t^{({\rm S})} \\ \dot{\E}_r^{({\rm S})} \\ \dot{\E}_T^{({\rm S})}\eem+
	\bem
	0 & 0 & 0 \\
	0 & 0 & 0 \\
	0 & 1 & 0
	\eem
	\bem \E_t^{({\rm S})}{}' \\ \E_r^{({\rm S})}{}' \\ \E_T^{({\rm S})}{}' \eem+
	\bem
	\f{k_{\rm S}}{r} & 0 & 0 \\
	0 & \f{k_{\rm S}}{r} & 0 \\
	0 & \f{1}{r} & \f{k_{\rm S}}{r}
	\eem
	\bem \E_t^{({\rm S})} \\ \E_r^{({\rm S})} \\ \E_T^{({\rm S})} \eem=\text{(sum of the other EOMs)}.
	\ee
Thus, one can recover $\E_t^{({\rm S})}$ and $\E_r^{({\rm S})}$ from the first- and second-line equations, and then $\E_T^{({\rm S})}$ can be written in terms of the other EOMs by use of the third-line equation.

\item $f_t^{({\rm S})}=H_L^{({\rm S})}=\delta\phi^{({\rm S})}=0$: \\
Since $f_t^{({\rm S})}$ cannot be defined for the monopole ($\ell=0$) mode, we focus on modes with $\ell\ge1$.
Note that $k_{\rm S}\ne0$.
Setting $\Delta_\e f_t^{({\rm S})}=\Delta_\e H_L^{({\rm S})}=\Delta_\e \delta\phi^{({\rm S})}=0$ in \eqref{gtsca}, we obtain
	\be
	\bem
	0 & -1 & 0 \\
	0 & 0 & 0 \\
	0 & 0 & 0
	\eem
	\bem \dot{T}_t^{({\rm S})} \\ \dot{L}^{({\rm S})} \\ \dot{T}_r^{({\rm S})}  \eem+
	\bem
	\f{k_{\rm S}}{r}  & 0 & 0 \\
	0 & -\f{k_{\rm S}}{(D-2)r} & -\f{B}{r} \\
	0 & 0 & -B\phi'
	\eem
	\bem T_t^{({\rm S})} \\ L^{({\rm S})} \\ T_r^{({\rm S})} \eem=
	\bem 0 \\ 0 \\ 0 \eem. \label{ex4}
	\ee
This satisfies the sufficient condition~\eqref{suf2}, and hence this gauge fixing is complete.
Indeed, the corresponding EOMs~$\E_t^{({\rm S})},\E_L^{({\rm S})}$ and $\E_{\delta\phi}^{({\rm S})}$ can be recovered from the Noether identity~\eqref{noetsca}:
	\be
	\bem
	0 & 0 & 0 \\
	1 & 0 & 0 \\
	0 & 0 & 0
	\eem
	\bem \dot{\E}_t^{({\rm S})} \\ \dot{\E}_L^{({\rm S})} \\ \dot{\E}_{\delta\phi}^{({\rm S})}\eem+
	\bem
	\f{k_{\rm S}}{r} & 0 & 0 \\
	0 & -\f{k_{\rm S}}{(D-2)r} & 0 \\
	0 & -\f{B}{r} & -B\phi' 
	\eem
	\bem \E_t^{({\rm S})} \\ \E_L^{({\rm S})} \\ \E_{\delta\phi}^{({\rm S})} \eem=\text{(sum of the other EOMs)}.
	\ee

\end{itemize}

\subsection{Unitary gauge}
\label{unitary}

In the context of scalar-tensor theories of gravity, the so-called unitary gauge is often used since it significantly simplifies the action.
In this gauge, one redefines time coordinate so that $\phi=t$, while spatial coordinates remain arbitrary.
Note that the unitary gauge fixing is valid in situations such as inflation or cosmology, where the gradient of the scalar field is timelike and the scalar field is monotonic in time.
For the following analysis we assume these conditions are satisfied.
 
Let us start from the general action~\eqref{generalST} and consider an infinitesimal transformation of time $t\to t+\e^0(x^\mu)$.
The corresponding gauge transformation of the scalar field is given by
	\be
	\Delta_\e\phi=-\dot \phi\,\e^0,
	\ee
and $\Delta_\e\phi=0$ has the unique solution $\e^0=0$.
This means that, starting from a configuration that satisfies the unitary gauge condition $\phi=t$, any infinitesimal gauge transformation spoils the gauge condition.
Thus, the unitary gauge fixing is complete and it can be imposed at the action level without losing any independent EOM by virtue of the main theorem.

This can be explicitly seen as follows. Let us denote the EOMs for $g_{\mu\nu}$ and $\phi$ as
	\be
	\E^{\mu\nu}\equiv\f{1}{\sqrt{-g}}\f{\delta S}{\delta g_{\mu\nu}},~~~\E_\phi\equiv\f{1}{\sqrt{-g}}\f{\delta S}{\delta \phi}.
	\ee
When one imposes the unitary gauge in the action, one does not obtain the scalar EOM $\E_\phi$ from the Euler-Lagrange equations.
Whether the lost EOM is redundant or not can be judged by use of the Noether identity. 
Since the action is invariant under the gauge transformation~\eqref{gencov}, we have
	\begin{align}
	0=\Delta_\e S&=\int d^Dx\sqrt{-g}\left[\E^{\mu\nu}(-2\nabla_\mu\e_\nu)+\E_\phi(-\e_\nu\nabla^\nu\phi)\right] \nonumber \\
	&=\int d^Dx\sqrt{-g}\left[2\nabla_\mu\E^{\mu\nu}-\E_\phi\nabla^\nu\phi\right]\e_\nu,
	\end{align}
so the Noether identity can be read off as
	\be
	2\nabla_\mu\E^{\mu\nu}-\E_\phi\nabla^\nu\phi=0.
	\ee
Hence, one finds $\E_\phi=(2/\dot{\phi})\nabla_\mu\E^\mu{}_0$ and thus the scalar EOM is redundant, as was stated in, e.g., \cite{Horndeski:1974wa}.

The unitary gauge has been employed to analyze complicated theories such as beyond Horndeski~\cite{Zumalacarregui:2013pma,Gleyzes:2014dya,Gao:2014soa}, whose action involves higher-order derivative terms.
It was shown in \cite{Gleyzes:2014dya,Gleyzes:2014qga,Lin:2014jga,Gao:2014fra,Saitou:2016lvb} that the class has 3 DOFs by Hamiltonian analyses in the unitary gauge.
However, \cite{Deffayet:2015qwa} pointed out that such analyses may not be appropriate since higher derivative terms in the action, which may yield Ostrogradsky ghost~\cite{Woodard:2015zca}, could be lost by the unitary gauge fixing (for the detailed arguments, see \cite{Blas:2009yd}).
Also, \cite{Langlois:2015cwa,Langlois:2015skt,Crisostomi:2016tcp} alerted the same problem.
Let us remark that these criticisms do not contradict the above argument based on our main theorem.
What we have shown is that the Lagrangian analysis does not change regardless of when one imposes the unitary gauge, i.e., before or after deriving EOMs.
Our work does not address the relation among DOFs in different gauges.

\section{Comments on recent works}
\label{comments}

From the discussions so far, we can draw a general lesson that a complete gauge fixing is harmless when deriving EOMs in the Lagrangian formalism, while an incomplete gauge fixing would result in some inconsistency.
Before making conclusions, let us revisit the confusions in recent works on counting DOFs in theories of modified gravity.

In the context of dRGT massive gravity, the dynamics of the St\"{u}ckelberg fluctuations around Minkowski background was investigated in \cite{Koyama:2011wx}.
Taking the so-called decoupling limit, St\"{u}ckelberg fluctuations can be decomposed as a sum of an additional St\"{u}ckelberg scalar $\pi$ and a free vector field $A^\mu$ with canonical Maxwell kinetic term, for which $U(1)$~symmetry is restored by $\pi$.
In this limit, \cite{Koyama:2011wx} imposed the Lorenz gauge condition to $A^\mu$ at the action level.
As explained in Sec.~III C of \cite{Motloch:2014nwa}, this process causes the problematic term $\dot\pi\dot A^0$, and leads to the inconsistent counting of DOFs.
It is now clear that the process confuses the DOF counting, as the Lorenz gauge fixing is not complete.
As we saw in \S \ref{ssec:bou}, if one imposes additional conditions $A^\mu\to 0$ as $|\vx|\to\infty$ and $A^0=0$, the Lorenz gauge fixing becomes complete.
Indeed, with these additional conditions, the problematic term vanishes.

Furthermore, a class of isotropic self-accelerating solutions in massive (bi)gravity was constructed in \cite{Gratia:2012wt,Motohashi:2012jd,Gratia:2013uza}.
They derived the EOMs for the St\"{u}ckelberg fields from the action with the metric ansatz~\eqref{gtssst} with $C(t,r)=0$ and $E(t,r)=1/B(t,r)$, derived the EOMs for the metric from the original action without the metric ansatz, and then imposed the metric ansatz to the EOMs for the metric.
While this metric ansatz itself is not a complete gauge fixing as we saw in \S \ref{ssec:sss}, the above process is consistent and does not lose any EOM since all the EOMs for the metric are derived consistently.  
It is interesting that the number of propagating DOFs for perturbations could still change.
Indeed, \cite{Khosravi:2013axa} showed that one of the kinetic terms for isotropic perturbations around the class of self-accelerating background~\cite{Gratia:2012wt} vanishes in some choice of coordinate.
It is clarified in \cite{Motloch:2015gta} that a poor choice of the coordinate in which the constant-time surface coincides with the characteristics of isotropic perturbations confuses the number of physical DOFs.
Although the situation is similar to the previous example, in this case the change of the number of physical DOFs is not originated from the issue of gauge fixing at the action level.

\section{Conclusions}
\label{conclusions}

Despite the long history of gauge theories and their analyses with gauge fixing, it had not been clarified under which condition gauge fixing at the action level is justified in the Lagrangian formalism, which caused some confusions in recent works.
Although the justification in the Hamiltonian formalism given in \cite{henneaux1992quantization} may also imply the validity in the Lagrangian formalism, it is still important to check this point explicitly to build a bridge to the practical implementation of gauge fixing like the ones in \S \ref{sec:app}.
In this paper, we addressed the issue under a general setting of gauge theory with multiple fields and multiple gauge symmetries defined in $D$-dimensional spacetime.
We proved the Theorem in \S \ref{sec:fth} that gauge fixing in the action yields consistent results if the gauge fixing is complete.
Our proof relies on the equivalence between the following two sets of EOMs:~one is the Euler-Lagrange equations derived from the original action supplemented with the gauge-fixing conditions, and the other is the Euler-Lagrange equations derived from the gauge-fixed action with Lagrange multipliers.
We showed that these two sets of EOMs coincide if the gauge fixing is complete, which is the consequence of the fact that the unique solvability is shared by a PDAE system and its adjoint system.

To apply the Theorem, one needs to check whether the gauge fixing of interest is complete or not.  
While it is not clear for general gauge theories how to check it, it is immediate to derive the necessary and sufficient condition~\eqref{mg1cond} for the case of $m_g=1$, i.e., field theories with a single gauge symmetry.  
Another possible case explored in \S \ref{sec:anm} is the case of $D=1$, which applies to analytical mechanics in arbitrary dimensions, or multiple fields with a homogeneous configuration, multiple fields in Fourier space, etc.  
We presented the sufficient conditions~\eqref{suf1} and \eqref{suf2}, and the necessary condition~\eqref{MN} for a gauge fixing to be complete, as well as the general methodology to judge whether a given gauge fixing is complete or not, for which the mathematical technique for ODAE systems is explained in Appendix~\ref{DAE}.
The examples provided in \S \ref{sec:app} illustrate applications of the above results.
They are helpful to resolve some of the confusions in recent papers as we commented in \S \ref{comments}.

While imposing gauge fixing at the action level is a powerful tool for analysis of gauge theories, it requires a special care as it may lead to some inconsistent result.
Our results elucidate that such a process leads to the same conclusion as the one obtained by imposing the gauge conditions after deriving EOMs if the gauge fixing is complete, and enable one to check whether or not the gauge fixing of interest is complete.

\acknowledgements{
We thank Stanley Deser, Wayne Hu, and Kiyoomi Kataoka for useful discussions.  
This work was supported in part by 
Japan Society for the Promotion of Science (JSPS) Grant-in-Aid for Young Scientists (B) No.\ 15K17632 (T.S.), 
MEXT Grant-in-Aid for Scientific Research on Innovative Areas ``New Developments in Astrophysics Through Multi-Messenger Observations of Gravitational Wave Sources'' No.\ 15H00777 (T.S.) and ``Cosmic Acceleration'' No.\ 15H05888 (T.S.).
}

\appendix

\section{Unique solvability of ODAEs}
\label{DAE}

\subsection{Standard canonical form of ODAEs}
\label{DAE1}

In this appendix, we discuss conditions for a system of ODAEs to have a unique solution without integration constant.
Since any higher-order DAE system can be recast into a first-order system by introducing auxiliary variables (see \S \ref{ssec:crit}), we consider a first-order ODAE system of the form
	\be
	M^I_J\dot{x}^J + N^I_Jx^J =g^I, \label{originalDAE}
	\ee
where $J=1,\cdots, m$.
In general, $I$ does not necessarily run over the same range as $J$, but in that case the system is obviously not uniquely solvable.
Hence, we assume $I$ also runs from $1$ to $m$.
The system is said to be uniquely solvable if and only if the following conditions are satisfied:
\begin{itemize}
\item The system is well posed, i.e., it has a solution for any inhomogeneity $g^I$.
\item No ODE appears.
\end{itemize}
It is obvious that the second requirement cannot be met if $\det M^I_J\ne0$.
We thus obtain a necessary (but not sufficient) condition for the unique solvability of the system:
	\be
	\det M^I_J=0.
	\ee
On the other hand, we can also derive the sufficient conditions~\eqref{suf1} and \eqref{suf2}.

A necessary and sufficient condition for the unique solvability is rather nontrivial~\cite{Berger:2013a}.
The idea is to recast the pair~$(M^I_J,N^I_J)$ into the form of \eqref{suf2} by transformation of variables with some regular matrices $S^I_J$ and $T^I_J$.
Let us multiply $S^I_J$ by both sides of \eqref{originalDAE} and write
	\be
	S^I_KM^K_LT^L_J\dot{\ti{x}}^J+(S^I_KN^K_LT^L_J+S^I_KM^K_L\dot{T}^L_J)\ti{x}^J=\ti{g}^I, \label{DAEtoSCF}
	\ee
where $\ti{x}^I$ and $\ti{g}^I$ are defined as
	\be
	\ti{x}^I\equiv (T^{-1})^I_Jx^J,~~~\ti{g}^I\equiv S^I_Jg^J.
	\ee
Therefore, if we define $\ti{M}^I_J$ and $\ti{N}^I_J$ by
	\be
	\ti{M}^I_J\equiv S^I_KM^K_LT^L_J,~~~\ti{N}^I_J\equiv S^I_KN^K_LT^L_J+S^I_KM^K_L\dot{T}^L_J,
	\ee
\eqref{DAEtoSCF} becomes 
	\be
	\ti{M}^I_J\dot{\ti{x}}^J+\ti{N}^I_J\ti{x}^J=\ti{g}^I, \label{DAEofSCF}
	\ee
which has the same form as \eqref{originalDAE}.
It has been shown in \cite{Berger:2013a} that, one can always choose $S^I_J$ and $T^I_J$ so that the pair of matrices $(\ti{M}^I_J,\ti{N}^I_J)$ takes the following ``standard canonical form'' (SCF):
	\be
	(\ti{M},\ti{N})=\left(
	\begin{bmatrix}
	I_{m_1}&0\\
	0&K_{m_2}(t)
	\end{bmatrix} ,
	\begin{bmatrix}
	J_{m_1}(t)&0\\
	0&I_{m_2}
	\end{bmatrix}
	\right) ,\label{SCF}
	\ee
if and only if the system is well-posed.
Here, $I_{m_i}$ denotes an $m_i\times m_i$ identity matrix, $K_{m_2}(t)$ is an $m_2\times m_2$ matrix which is strictly lower triangular, $J_{m_1}(t)$ is some $m_1\times m_1$ matrix, and $m_1+m_2=m$.
From the block-diagonal structure of \eqref{SCF}, it is clear that in \eqref{DAEofSCF} the equations for the first $m_1$ variables and the last $m_2$ are decoupled.  
If $m_1\neq 0$, since the upper-left $m_1\times m_1$ submatrix of $\ti{M}^I_J$ is the identity matrix, the first $m_1$ equations are inevitably ODEs and thus the unique solvability of the system is spoiled.  If $m_1=0$, one is left with equations of the form
	\be \label{soldae}
	\begin{bmatrix}
	0&\cdots&\cdots&0 \\
	\ast&\rotatebox{-5}{$\ddots$}&&\vdots \\
	\vdots&\rotatebox{-5}{$\ddots$}&\rotatebox{-5}{$\ddots$}&\vdots \\
	\ast&\cdots&\ast&0
	\end{bmatrix}
	\begin{bmatrix}
	\dot{\ti{x}}^1\\
	\vdots \\
	\vdots \\
	\dot{\ti{x}}^m
	\end{bmatrix}+
	\begin{bmatrix}
	\ti{x}^1\\
	\vdots \\
	\vdots \\
	\ti{x}^m
	\end{bmatrix}=
	\begin{bmatrix}
	\ti{g}^1\\
	\vdots \\
	\vdots \\
	\ti{g}^m
	\end{bmatrix}.
	\ee
This precisely satisfies the sufficient condition~\eqref{suf2}.
We can uniquely solve this ODAE system for $\ti{x}^I$ from the first-line equation to the $m$th-line equation without any integration constant,
and then obtain $x^I$ through $x^I=T^I_J \ti{x}^J$.

In conclusion, the necessary and sufficient condition for the unique solvability of the DAE system~\eqref{originalDAE} with a matrix pair $(M,N)$ is that the corresponding SCF~\eqref{SCF} has $m_1=0$, namely, 
	\be \label{SCFsol} (\ti{M},\ti{N})=(K_m,I_m). \ee
Obviously, for a vanishing source term $g^I$, the unique solution is given by $x^I=0$.

As an application of the above methodology, let us consider the following ODAE system:
	\be
	\bem
	2 & t+2 & -t-1 \\
	-2t & -t(t+2) & t(t+1) \\
	2t & t(t+1) & -t^2
	\eem
	\bem \dot{x}^1 \\ \dot{x}^2 \\ \dot{x}^3 \eem+
	\bem
	1 & -t+1 & t \\
	-t+2 & t^2+1 & -t(t+1) \\
	0 & t+1 & -t-1
	\eem
	\bem x^1 \\ x^2 \\ x^3 \eem=
	\bem 0 \\ 0 \\ 0 \eem, \label{scfex}
	\ee
which corresponds to the case of
	\be
	(M,N)=\left(\bem
	2 & t+2 & -t-1 \\
	-2t & -t(t+2) & t(t+1) \\
	2t & t(t+1) & -t^2
	\eem,
	\bem
	1 & -t+1 & t \\
	-t+2 & t^2+1 & -t(t+1) \\
	0 & t+1 & -t-1
	\eem\right),~~~g^I=0.
	\ee
In this case, we can find the regular transformation matrices $S,T$ as
	\be
	S=\bem
	t & 1 & 0 \\
	0 & 0 & 1 \\
	1 & 0 & 0
	\eem,~~~
	T=\bem
	1 & -2t-1 & -1 \\
	-1 & 3t+2 & 2 \\
	-1 & 3t+1 & 2
	\eem.
	\ee
These matrices actually transform the pair $(M,N)$ into the SCF:
	\be
	(\ti{M},\ti{N})=\left(\bem
	0 & 0 & 0 \\
	t & 0 & 0 \\
	1 & 1 & 0
	\eem,
	\bem
	1 & 0 & 0 \\
	0 & 1 & 0 \\
	0 & 0 & 1
	\eem\right),
	\ee
which satisfies the sufficient condition~\eqref{suf2}.
This means that the system~\eqref{scfex} is uniquely solvable and the solution is given by $x^I=0$.

\subsection{Adjoint ODAE}
\label{DAE2}

As we proved in \S \ref{sec:mth}, if a DAE system is uniquely solvable, its adjoint DAE system is also uniquely solvable.
Here, we show the fact for ODAEs in a more direct manner.
The adjoint ODAE system to \eqref{originalDAE} has the form of
	\be
	\f{d}{dt}(y_JM^J_I)-y_JN^J_I=h_I. \label{adjDAE}
	\ee
If the ODAE system~\eqref{originalDAE} is uniquely solvable, there exists a pair of matrices $(S,T)$ that transforms $(M,N)$ into 
the form of \eqref{SCFsol}.
Using the pair $(S,T)$, we can rewrite \eqref{adjDAE} as
	\be
	\f{d}{dt}(\ti{y}_J\ti{M}^J_I)-\ti{y}_J\ti{N}^J_I=\ti{h}_I, \label{adjDAEofSCF}
	\ee
where we have defined
	\be
	\ti{y}_I\equiv y_J(S^{-1})^J_I,~~~\ti{h}_I\equiv h_JT^J_I.
	\ee
More explicitly, \eqref{adjDAEofSCF} can be written as
	\be
	{}^t\!K_m {}^t\!\dot{\ti{y}} - (I_m-{}^t\!\dot{K}_m) {}^t\!\ti{y}={}^t\!\ti{h},
	\ee
namely,
	\be
	\begin{bmatrix}
	0&\ast&\cdots&\ast \\
	\vdots&\rotatebox{-5}{$\ddots$}&\rotatebox{-5}{$\ddots$}&\vdots \\
	\vdots&&\rotatebox{-5}{$\ddots$}&\ast \\
	0&\cdots&\cdots&0
	\end{bmatrix}
	\begin{bmatrix}
	\dot{\ti{y}}_1\\
	\vdots \\
	\vdots \\
	\dot{\ti{y}}_m
	\end{bmatrix}+
	\begin{bmatrix}
	1&\ast&\cdots&\ast \\
	0&\rotatebox{-5}{$\ddots$}&\rotatebox{-5}{$\ddots$}&\vdots \\
	\vdots&\rotatebox{-5}{$\ddots$}&\rotatebox{-5}{$\ddots$}&\ast \\
	0&\cdots&0&1
	\end{bmatrix}
	\begin{bmatrix}
	\ti{y}_1\\
	\vdots \\
	\vdots \\
	\ti{y}_m
	\end{bmatrix}=
	\begin{bmatrix}
	\ti{h}_1\\
	\vdots \\
	\vdots \\
	\ti{h}_m
	\end{bmatrix}.
	\ee
Similarly to \eqref{soldae}, this system satisfies the sufficient condition~\eqref{suf2}. It can be solved for $\ti{y}_I$ from the $m$th-line equation to the first-line equation without any integration constant.  In particular, for a homogeneous system with $h_I=0$, the unique solution is $y_I=0$.

\section{Harmonic functions}
\label{hf}

In this appendix, we briefly summarize the definitions of the harmonic tensors, vectors and scalars in general spacetime dimension~\cite{Kodama:2000fa}.
In what follows, $\gamma_{ij}$ represents the metric of an $n$-dimensional constant-curvature space with $n\ge 2$, and $D_i$ denotes a covariant derivative with respect to $\gamma_{ij}$.
Here, we restrict ourselves to the case of spatial curvature~$\kappa=1$.
For a construction of the harmonic functions and their eigenvalues and degeneracies, see \cite{Rubin:1984}.

\subsection{Tensor}
The harmonic tensors $\bT_{ij}$ are defined so that they satisfy
	\be
	(\Lap +k_{\rm T}^2)\bT_{ij}=0,~~~\bT_{ij}=\bT_{ji},~~~\bT^i_i=0,~~~D_j\bT^j_i=0,
	\ee
where $\Lap\equiv D^iD_i$.
Note that $\bT_{ij}$ becomes trivial in two-dimensional space.
The eigenvalue $k_{\rm T}^2$ takes discrete values~\cite{Rubin:1984}
	\be
	k_{\rm T}^2=\ell(\ell+n-1)-2,~~~(\ell=2,3,\cdots),
	\ee
and hence always positive.

\subsection{Vector}
The harmonic vectors $\bV_i$ are defined by
	\be
	(\Lap +k_{\rm V}^2)\bV_i=0,~~~D_i\bV^i=0,
	\ee
where the eigenvalue $k_{\rm V}^2$ is given by~\cite{Rubin:1984}
	\be
	k_{\rm V}^2=\ell(\ell+n-1)-1,~~~(\ell=1,2,\cdots),
	\ee
and all positive.
One can construct the vector-type harmonic tensor from the vector harmonic function $\bV_i$ as
	\be
	\bV_{ij}=-\f{1}{2k_{\rm V}}(D_i\bV_j+D_j\bV_i),
	\ee
which satisfies
	\be
	\left[ \Lap+k_{\rm V}^2-(n+1)\right]\bV_{ij}=0,~~~\bV^i_i=0,~~~D_j\bV^j_i=\f{k_{\rm V}^2-(n-1)}{2k_{\rm V}}\bV_i.
	\ee
For $\bV_{ij}$ to be nonvanishing, it is necessary that $k_{\rm V}^2>n+1$ as the operator $\Lap$ is negative definite.\footnote{This means that, for any function $H$ that satisfies $(\Lap+k^2)H=0$, the eigenvalue $-k^2$ is negative, i.e., $k^2>0$.}
Therefore, $\bV_{ij}$ becomes nontrivial only for $\ell\ge 2$.

\subsection{Scalar}
The scalar harmonic functions $\bS$ are defined by
	\be
	(\Lap +k_{\rm S}^2)\bS=0,
	\ee
where the eigenvalue $k_{\rm S}^2$ takes~\cite{Rubin:1984}
	\be
	k_{\rm S}^2=\ell(\ell+n-1),~~~(\ell=0,1,\cdots). \label{seigen}
	\ee
From the scalar harmonic function $\bS$, we can construct the scalar-type harmonic vector $\bS_i$ as
	\be
	\bS_i=-\f{1}{k_{\rm S}}D_i\bS,
	\ee
which has the properties
	\be
	\left[ \Lap+k_{\rm S}^2-(n-1)\right]\bS_i=0,~~~D_i\bS^i=k_{\rm S}\bS.
	\ee
The scalar-type harmonic tensor $\bS_{ij}$ is defined by
	\be
	\bS_{ij}=\f{1}{k_{\rm S}^2}D_iD_j\bS+\f{1}{n}\gamma_{ij}\bS,
	\ee
which satisfies
	\be
	\left( \Lap+k_{\rm S}^2-2n\right)\bS_{ij}=0,~~~\bS^i_i=0,~~~D_j\bS^j_i=\f{n-1}{n}\f{k_{\rm S}^2-n}{k_{\rm S}}\bS_i.
	\ee
Note that $k_{\rm S}^2=0$ for $\ell=0$, so we cannot define $\bS_i$ or $\bS_{ij}$.
$\bS_i$ is properly defined for $\ell\ge 1$, since one obtains $k_{\rm S}^2>n-1$ for these modes.
As for $\bS_{ij}$, $k_{\rm S}^2>2n$ requires $\ell\ge 2$.


\bibliography{gauge-fix}

\end{document}